# Rechargeable Li/Cl$_2$ battery down to -80 °C

Peng Liang, Guanzhou Zhu, Cheng-Liang Huang, Yuan-Yao Li, Hao Sun, Bin Yuan, Shu-Chi Wu, Jiachen Li, Feifei Wang, Bing-Joe Hwang, and Hongjie Dai[*]

P. Liang, G. Zhu, S.-C. Wu, J. Li, H. Dai
Department of Chemistry and Bio-X, Stanford University, Stanford, CA 94305, USA
E-mail: hdai1@stanford.edu

C.-L. Huang, Y.-Y. Li
Department of Chemical Engineering, National Chung Cheng University, Chia-Yi, 62102, Taiwan

C.-L. Huang
Department of Electrical Engineering, National Chung Cheng University, Chia-Yi, 62102, Taiwan

H. Sun, B. Yuan
Frontiers Science Center for Transformative Molecules, School of Chemistry and Chemical Engineering, and Zhangjiang Institute for Advanced Study, Shanghai Jiao Tong University, Shanghai, 200240, China

F. Wang
Department of Electrical and Electronic Engineering, The University of Hong Kong, Hong Kong 999077, Hong Kong

B.-J. Hwang
Department of Chemical Engineering, National Taiwan University of Science and Technology, Taipei, 106, Taiwan

**Abstract**

Low temperature rechargeable batteries are important to life in cold climates, polar/deep-sea expeditions and space explorations. Here, we report ~ 3.5 - 4 V rechargeable lithium/chlorine (Li/Cl$_2$) batteries operating down to -80 °C, employing Li metal negative electrode, a novel CO$_2$ activated porous carbon (KJCO$_2$) as the positive electrode, and a high ionic conductivity (~ 5 to 20 mS cm$^{-1}$ from -80 °C to 25 °C) electrolyte comprised of 1 M aluminum chloride (AlCl$_3$), 0.95 M lithium chloride (LiCl), and 0.05 M lithium bis(fluorosulfonyl)imide (LiFSI) in low melting point (-104.5 °C) thionyl chloride (SOCl$_2$). Between room-temperature and -80 °C, the Li/Cl$_2$ battery delivered up to ~ 30,000 - 4,500 mAh g$^{-1}$ first discharge capacity and a 1,200 - 5,000 mAh g$^{-1}$ reversible capacity (discharge voltages in ~ 3.5 to 3.1 V) over up to 130 charge-discharge cycles. Mass spectrometry and X-ray photoelectron spectroscopy (XPS) probed Cl$_2$ trapped in the porous carbon upon LiCl electro-oxidation during charging. At lower temperature down to -80 °C, SCl$_2$/S$_2$Cl$_2$ and Cl$_2$ generated by electro-oxidation in the charging step were trapped in porous KJCO$_2$ carbon, allowing for reversible reduction to afford a high discharge voltage plateau near ~ 4 V with up to ~ 1000 mAh g$^{-1}$ capacity for SCl$_2$/S$_2$Cl$_2$ reduction and up to ~ 4000 mAh g$^{-1}$ capacity at ~ 3.1 V plateau for Cl$_2$ reduction. Towards practical use, we made CR2032 Li/Cl$_2$ battery cells to drive digital watches at -40 °C and light emitting diode at -80 °C, opening Li/Cl$_2$ secondary batteries for ultra-cold conditions.

**1. Introduction**

Existing rechargeable batteries underperform at low temperatures (< ~ -30 °C) with reduction in battery capacity and cycle-life, limiting the operation of electronic devices, electric vehicles and equipment in polar climates, subsea and space explorations[1-6]. Li-ion batteries with ethylene carbonate (EC)-based electrolytes can function down to ~ -20 °C[7], limited by decreased ionic conductivity, slower interfacial charge transfer kinetics and Li$^+$ transport in electrodes[8-14]. Developing new rechargeable batteries at low temperatures down to -80 °C has been challenging[15-19], especially for batteries with high specific capacity and energy density.

The lithium-thionyl chloride (Li-SOCl$_2$) primary batteries are well known for their high energy density and wide use in professional electronics and industrial instruments, but lacks rechargeability[20, 21]. Although at room temperature (RT) we converted Li-SOCl$_2$ primary batteries into Na/Cl$_2$ (or Li/Cl$_2$) secondary batteries via mainly redox between Cl$_2$/Cl$^-$ at the positive electrode and Na/Na$^+$ (or Li/Li$^+$) redox at the negative electrode[20, 21], thus far rechargeable Li/Cl$_2$ batteries remains unexplored below room temperature (RT) due to the limitation of electrolyte and positive electrode material[22-25]. More importantly, the Li/Cl$_2$ battery chemistry at low temperature is largely unknow.

Here, we investigate an ultra-low temperature Li/Cl$_2$ battery using a Li metal as the negative electrode, Ketjenblack (KJ) carbon black activated in carbon dioxide (CO$_2$) at 1000 °C for 45 min (KJCO$_2$, see Methods) as the positive electrode, and AlCl$_3$/LiCl/LiFSI/dissolved in SOCl$_2$ as the electrolyte. The newly engineered electrolyte affords low viscosity and high ionic conductivity at low temperatures (5.26 mS cm$^{-1}$ at -80 °C) using an extreme low concentration of AlCl$_3$ (1.0 M), and stable passivation of the Li metal anode with a solid-electrolyte-interface (SEI) rich in LiCl/LiF. Also important is high temperature CO$_2$ activation of KJ carbon affording ~ 2X increase in surface area (~ 2386.9 m$^2$ g$^{-1}$) and pore volume (~ 6.5 cm$^3$ g$^{-1}$) for the positive electrode and giving much more LiCl deposition during the first discharge for subsequent reversible LiCl/Cl$_2$ redox reactions, and delivering up to ~ 30,000 mAh g$^{-1}$ first discharge capacity (~3.55 V), which is the highest capacity known to date. The resulting batteries deliver a 1,200 - 5,000 mAh g$^{-1}$ reversible capacity operating between room-temperature (25 °C) and -80 °C. By X-ray and mass spectroscopy, we studied the battery reaction chemistry and analyzed the main reactive species generated by battery charging at -40 °C to -80 °C and trapped in porous carbon including SCl$_2$/S$_2$Cl$_2$, SOCl$_2$, SO$_2$Cl$_2$ and Cl$_2$, shedding light to battery rechargeability at low temperatures.

## 2. Results and Discussion
### 2.1. The first discharge of Li/KJCO$_2$ batteries

We constructed CR2032 coin cell (20 mm diameter x 3.2 mm thickness) batteries by using a Li metal as the negative electrode, either KJ (Li/KJ battery) or KJCO$_2$ (Li/KJCO$_2$ battery) as the positive electrode in an electrolyte comprised of 1.0 M AlCl$_3$ + 0.95 M LiCl dissolved in SOCl$_2$ with 0.05 M of LiFSI additive (Fig. 1a, see Methods). The as received KJ material (EC-600JD) (Fig. 1b)[20] showed signs of etching and apparent reduction in size after activation in CO$_2$ at 1000 °C for 45 minutes (Fig.1c), accompanied by a mass loss of ~ 85% due to the reaction CO$_2$ (g) + C (s) →2 CO (g)[20, 21]. Raman spectroscopy showed increased defects and disorder (Supplementary Fig. 1a)[26], with a complete disappearance of graphitic ordering in X-ray diffraction after the CO$_2$ activation step (Supplementary Fig. 1b)[27].

At RT the Li/KJCO$_2$ battery showed a 2X higher first discharge capacity than Li/KJ cell (~ 15,300 mAh g$^{-1}$ or 30.6 mAh cm$^{-2}$ and ~ 29,100 mAh g$^{-1}$ or 58.2 mAh cm$^{-2}$ for KJ and KJCO$_2$ respectively; current = 50 mA g$^{-1}$, Fig. 1d) with a ~ 3.55 V discharge voltage. Note that throughout this work, specific capacity and areal capacity were based on mass of loaded carbon in the positive electrode (~ 2 mg cm$^{-2}$) and electrode area (~ 1.767 cm$^2$) respectively, as commonly done for Li/air or Li/O$_2$ batteries[28]. Higher KJCO$_2$ mass loading of ~ 6 mg cm$^{-2}$ still afforded an impressive first discharge capacity of ~ 9,407 mAh g$^{-1}$ at the same current (Supplementary Fig. 2). The first discharge (~ 3.55 V plateau) was due to Li metal oxidation to Li$^+$ on the negative electrode and the SOCl$_2$ reduction to sulfur (S), sulfur dioxide (SO$_2$) and LiCl deposited on the porous carbon positive electrode (2 SOCl$_2$ + 4 Li$^+$ + 4 e$^-$ → S + SO$_2$ + 4 LiCl), accompanied by a discernible ~ 3.2 V plateau attributed to reduction of SO$_2$[20, 21, 29, 30]. The CO$_2$ activation process doubled the specific surface area (from 1307.4 m$^2$ g$^{-1}$ to 2386.9 m$^2$ g$^{-1}$) and pore volume for KJ to KJCO$_2$ (from 3.1 cm$^3$ g$^{-1}$ to 6.5 cm$^3$ g$^{-1}$) (Fig. 1e and Supplementary Table 1), allowing for increased deposition of LiCl and higher discharge capacity (Fig. 1d)[20, 31]. High surface area and large pore volume of carbon materials are also found important to the reversible specific capacity of secondary Na/Cl$_2$ batteries[20].

At low temperatures, the Li/KJCO$_2$ battery delivered first discharge voltage/capacity of 3.4 V/8,521 mAh g$^{-1}$, 3.3 V/5,532 mAh g$^{-1}$, and 3.1 V/4,503 mAh g$^{-1}$ at -20 °C, -40 °C and -80 °C, respectively (Fig. 2a). LiCl deposition on KJCO$_2$ was confirmed by XRD (Fig. 2b), both in the pores of KJCO$_2$ and on the surfaces of KJCO$_2$ until passivation (see schematic in Fig. 2c)[20, 21]. This was consistent with that pore volumes of carbon materials in the positive electrode positively correlated with the first discharge capacity of primary batteries[31-34] and in recent secondary batteries[20, 21]. Based on the SOCl$_2$ reduction reaction equation and the 1$^{st}$ discharge capacity, the estimated deposited LiCl volume was ~ 0.07 cm$^3$, ~ 0.027 cm$^3$, ~ 0.013 cm$^3$ and ~ 0.011 cm$^3$ at RT, -20 °C, -40 °C and -80 °C respectively compared to the total pore volume ~ 0.023 cm$^3$ in the KJCO$_2$ electrode.

LiCl deposition on KJCO$_2$ decreased with reduced discharge capacity at lower temperatures (Fig. 2a) due to increased electrolyte viscosity (Supplementary Fig. 3), decreased Li$^+$ and Cl$^-$ ion diffusion[22, 23, 35] and lower electrolyte conductivity from 20.34 mS cm$^{-1}$ at room temperature to 5.26 mS cm$^{-1}$ at -80 °C (Fig. 2d). Due to slower growth kinetics of crystallites at lower temperatures[22, 23, 35], the size of LiCl crystallites deposited on the electrode decreased from > 3 µm at room temperature (Fig. 2e) to ~ 1.5 µm, ~ 200 nm and ~ 60 nm at -20 °C, -40 °C and -80 °C respectively (Fig. 2f-h) accompanied by weaker XRD patterns (Fig. 2b).

The 1 M neutral electrolyte and low melting point (-104.5 °C) of SOCl$_2$ were important to battery operation down to -80 °C owing to low electrolyte viscosity than those in our previous alkali metal/Cl$_2$ batteries (with higher 4M or 1.8 M AlCl$_3$, Supplementary Fig. 4)[20-22]. Through CO$_2$ activation of a common carbon black material and electrolyte tuning, we obtained a record setting first discharge capacity up to ~ 30,000 mAh g$^{-1}$ with flat ~ 3.55 V discharge plateau, an important result on its own right in terms of high energy density Li/SOCl$_2$ primary batteries.

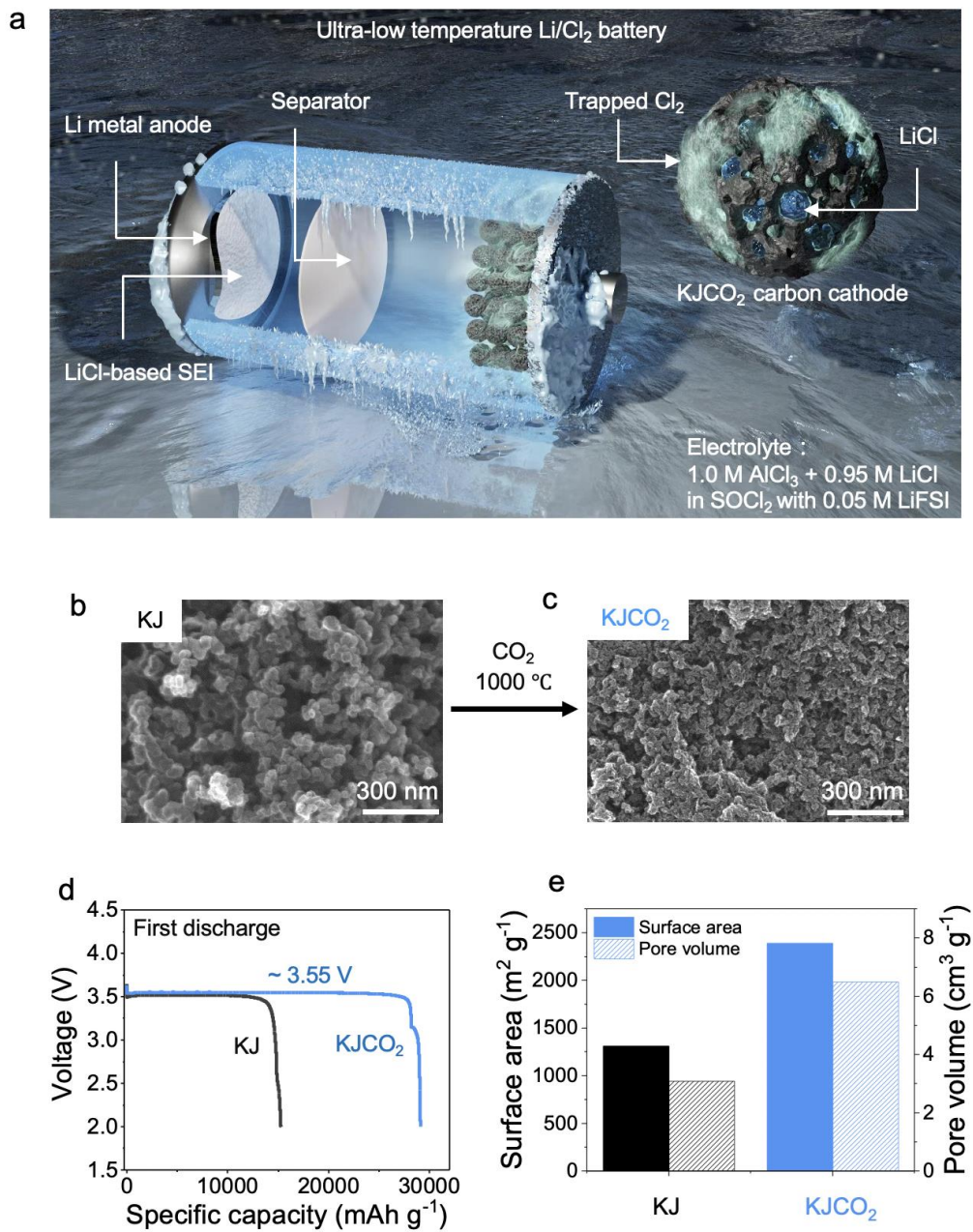

**Figure 1. Ultrahigh first discharge capacity of a Li/porous-carbon battery at room temperature. a**, Schematic drawing of an ultra-low temperature Li/Cl$_2$ battery. **b. c**, SEM images of pristine KJ (b) and high temperature activated KJCO$_2$ (c) at the same magnification. **d**, The first discharge curves of room temperature batteries using either KJ (black) or KJCO$_2$ (blue) recorded at 50 mA g$^{-1}$. Notice the drastic discharge capacity increase when KJCO$_2$ was the positive electrode. **e**, Average measured BET surface area, pore volume (micropores and mesopores) of pristine KJ (black) or KJCO$_2$ (blue). The loading of KJ or KJCO$_2$ was ~ 2.0 mg cm$^{-2}$ in the positive electrodes for the Li/KJ or Li/KJCO$_2$ batteries.

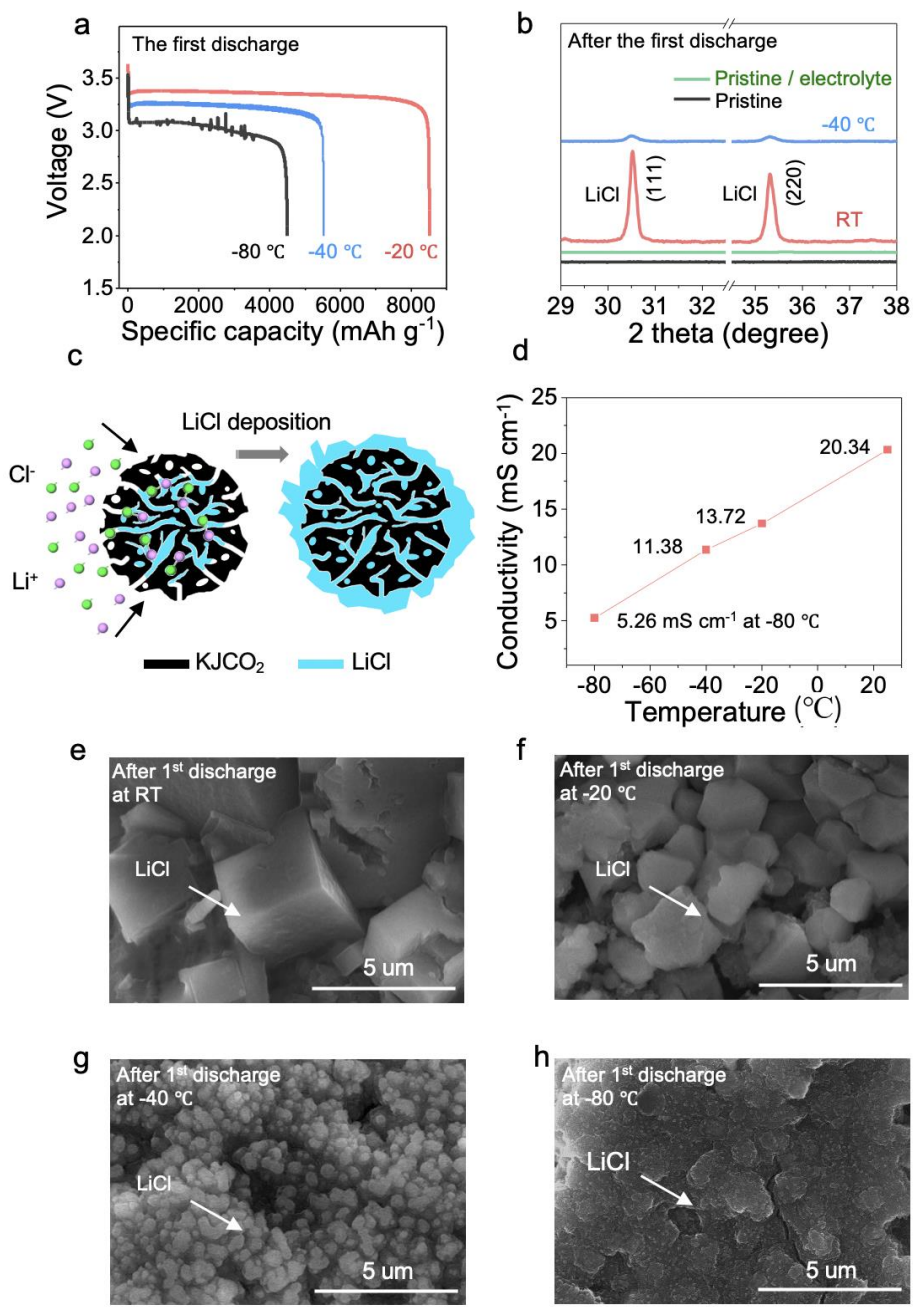

**Figure 2. The first discharge characteristics of Li/KJCO₂ batteries at low temperatures. a**, The first discharge curves of Li/KJCO₂ batteries at -20 °C, -40 °C and -80 °C respectively. The current was 50 and 12.5 mA g$^{-1}$ for -20/-40 °C and -80 °C, respectively. **b**, XRD of pristine KJCO₂ electrode (black), pristine KJCO₂ electrode simply soaked in electrolyte and dried (green, used as a control sample), and KJCO₂ electrode after the battery's first discharge at room temperature (red) and -40 °C (blue). LiCl was formed on the KJCO₂ electrode after first discharge, indicated by the strong LiCl XRD peaks. **c**. Schematic drawing of the LiCl deposition in the pores and on the surface of KJCO₂. **d**. The measured ionic conductivity of the electrolyte (1.0 M AlCl₃ + 0.95 M LiCl dissolved in SOCl₂ with 0.05 M of LiFSI additive) at different temperatures. **e-h**, SEM images of KJCO₂ electrode after the first discharge at RT, -20 °C, -40 °C and -80 °C, showing different morphologies/reduced sizes of the deposited LiCl. The loading of KJCO₂ was ~ 2.0 mg cm$^{-2}$ on the positive electrodes of all batteries.

**2.2. Secondary batteries operating at low temperatures at -20 °C, -40 °C and -80 °C.**

In the temperature range of RT to -20°C, after the first discharge the Li/KJCO$_2$ cell was rechargeable and cyclable at 1200 mAh g$^{-1}$ capacity (areal capacity ~ 2.4 mAh cm$^{-2}$, current 100 mA g$^{-1}$) with up to ~ 70 cycle-life (Supplementary Fig. 5). Rechargeability mainly stemmed from reversible LiCl/Cl$_2$ redox reactions on the positive electrode, i.e., Cl$_2$ was generated by LiCl oxidation during charging and trapped in the pores of KJCO$_2$, and reduced to LiCl upon discharge[20, 21]. Note that throughout this work, charging was controlled by setting the charging time at a specific capacity (= cycling capacity/current) and discharge proceeded down to a cut-off voltage of 2.0 V (see Methods).

At -40 °C the Li/KJCO$_2$ battery exhibited improved cycle life over ~ 130 cycles with 1200 mAh g$^{-1}$ capacity under 100 mA g$^{-1}$ current, with an average coulombic efficiency (CE) was ~ 100% (Fig. 3a). Further, the battery showed cyclability at a higher specific capacity of 2500 mAh g$^{-1}$ (areal capacity: ~ 5 mAh cm$^{-2}$ per cycle) at -40 °C over > 75 cycles (Fig. 3b), much improved over at -20 °C (< 15 cycles, Supplementary Fig. 6). Continued increasing of the cycling capacities to 4000 mAh g$^{-1}$ (areal capacity ~ 8 mAh cm$^{-2}$) and 5000 mAh g$^{-1}$ (areal capacity ~ 10 mAh cm$^{-2}$) still exhibited rechargeability/cyclability albeit with a reduced cycle-life of < 10 cycles (Fig. 3c, d). The battery's CE initially deviated from 100% (Fig. 3a-d), suggesting an "in situ activation" processes such as anionic intercalation/expanding/exfoliation of disordered carbon layers in the positive electrode as the battery started charging and cycling[21].

The high cycling capacity of our Li/KJCO$_2$ battery at -40 °C (1200 mAh g$^{-1}$, > 130 cycles; 2500 mAh g$^{-1}$, > 75 cycle) and ~ 3.3 V discharging plateau (Fig. 3e) encouraged us to employ a Li/KJCO$_2$ CR2032 coin cell charged to 5000 mAh g$^{-1}$ to power an electronic watch purchased from Amazon operating at a voltage of 3.0 V in a -40 °C freezer. The fully charged battery powered the watch at -40 °C over ~ 6 months (Fig. 3f) on one charging.

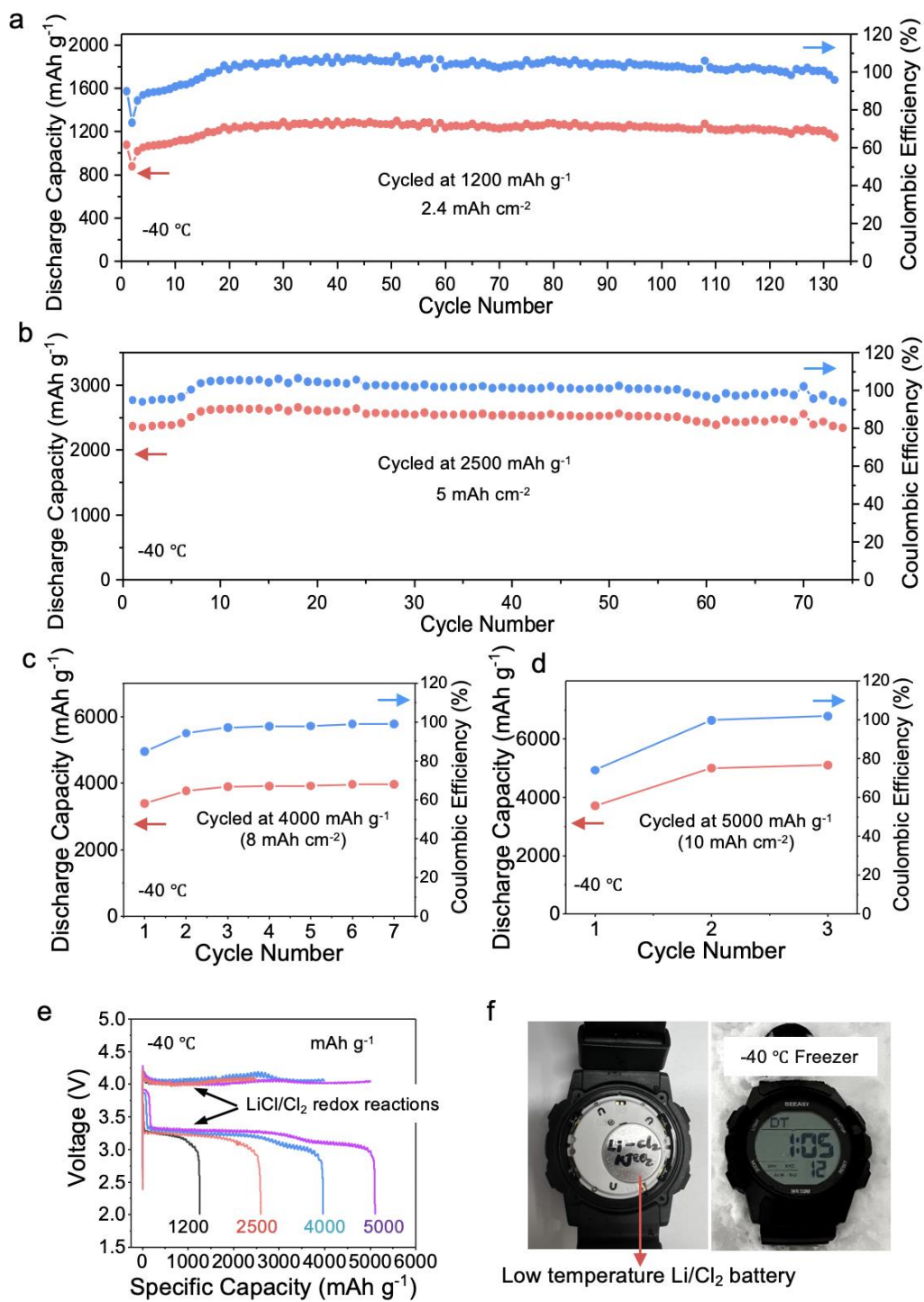

**Figure 3. The cycling characteristics of Li/KJCO$_2$ batteries at -40 °C. a-d**, Cycling performance of a Li/KJCO$_2$ battery (KJCO$_2$) at 1200 mAh g$^{-1}$ (a), 2500 mAh g$^{-1}$ (b), 4000 mAh g$^{-1}$ (c), and 5000 mAh g$^{-1}$ (d) cycling capacity under 100 mA g$^{-1}$ current at -40 °C. **e**, Charge-discharge curves of Li/KJCO$_2$ batteries at 1200 - 5000 mAh g$^{-1}$ cycling capacities under 100 mA g$^{-1}$ current at -40 °C. **f**, Photos showing a commercial electronic watch powered by a fully charged Li/KJCO$_2$ battery coin cell (CR2032) in a -40 °C freezer over 6 months. The loading of KJCO$_2$ was ~ 2.0 mg cm$^{-2}$ in all batteries in this figure.

At -80 °C, the Li/KJCO$_2$ batteries (placed in a biological freezer) cycled stably at 1200 mAh g$^{-1}$ over ~ 70 cycles with near ~100 % average CE (Fig. 4a, current 25 mA g$^{-1}$). At this rate batteries were also cyclable at higher specific capacity of 2500 mAh g$^{-1}$ (Fig. 4b), 4000 mAh g$^{-1}$ (Fig. 4c) and even ~ 5000 mAh g$^{-1}$ (Supplementary Fig. 7), but with shorter cycle lives. At -80 °C, the main battery charging plateau of ~ 4.25 V and discharging plateau of ~3.1 V discharging plateau were attributed to oxidation of LiCl to Cl$_2$ and the reduction of Cl$_2$ back to LiCl, respectively (Fig. 4d). In addition to LiCl oxidation on KJCO$_2$, charging to higher capacities also led to increased oxidation of SOCl$_2$ in the electrolyte to form SCl$_2$/S$_2$Cl$_2$, especially when charging to ~ 5000 mAh g$^{-1}$ during which a pronounced increase in charging voltage was observed (Fig. 4d). This was also consistent with the appearance of an impressive high discharge voltage plateau at ~ 3.8 V, with a specific capacity up to 1000 mAh g$^{-1}$ (Fig. 4d) due to reduction of the SCl$_2$/S$_2$Cl$_2$ generated in the charging step[20, 21, 29]. The discharge plateaus at ~ 3.1 V and ~ 2.9 V corresponded to reduction of Cl$_2$ and SO$_2$Cl$_2$, respectively (Fig. 4d, Supplementary Text 1)[20, 21].

We employed a battery placed in a -80 °C freezer (Fig. 4e-g) charged to ~ 5000 mAh g$^{-1}$ to power a light-emitting diode (LED) (with operating voltage > 3.0 V; Fig. 4f shows electrolyte remained as a liquid at -80 °C).

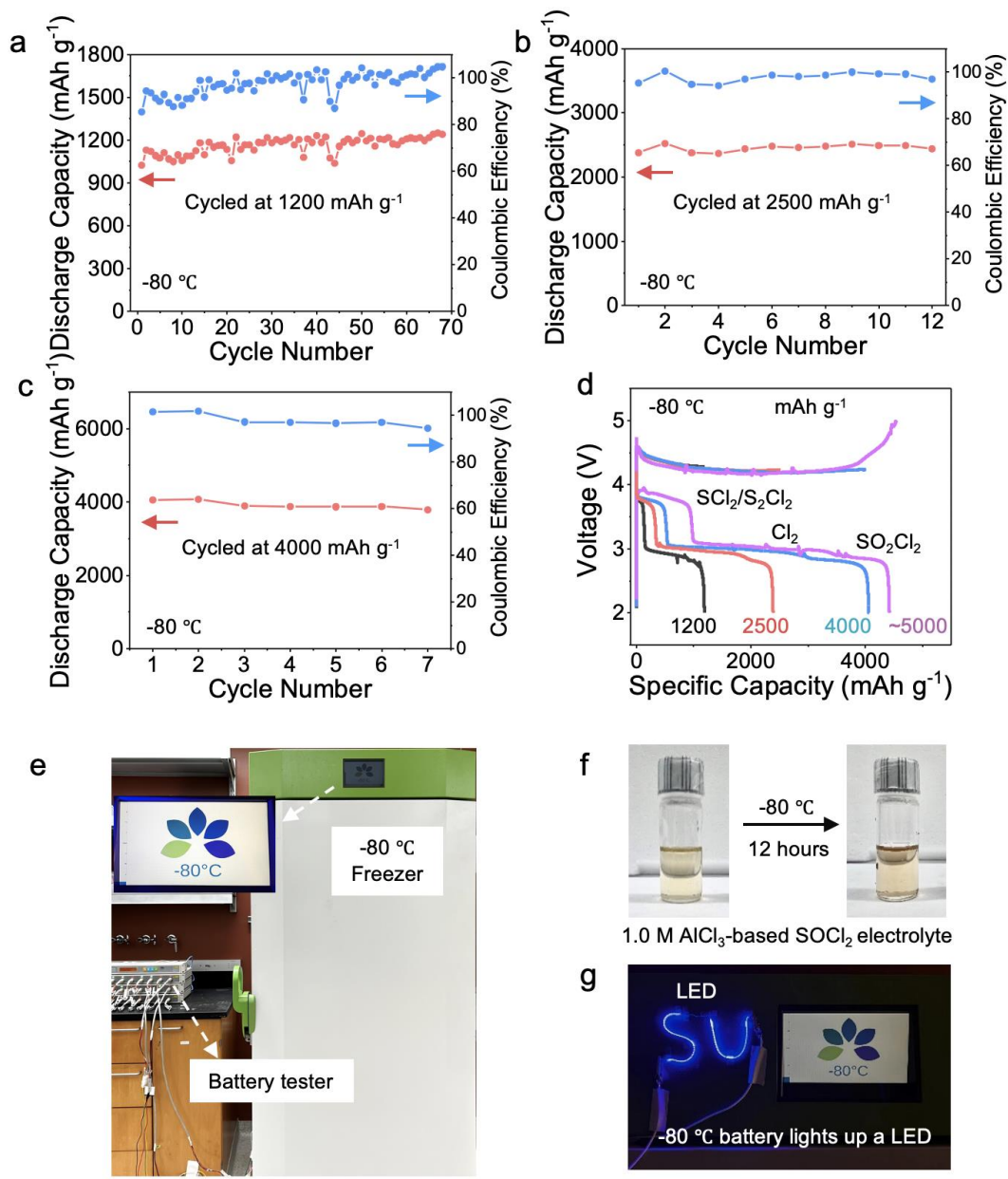

**Figure 4. The cycling characteristics of Li/KJCO₂ batteries at ultra-low temperature (-80 °C). a-c**, Cycling performance of a Li/KJCO₂ battery with 1200 mAh g⁻¹ (a), 2500 mAh g⁻¹ (b), and 4000 mAh g⁻¹ (c) cycling capacity under 25 mA g⁻¹ current at -80 °C. **d**, Charge-discharge curves of Li/KJCO₂ batteries with 1200 - ~ 5000 mAh g⁻¹ cycling capacities under 25 mA g⁻¹ current at -80 °C (Note that the voltage limit of battery tester was 5.0 V). **e**, The -80 °C freezer and battery testers used for low temperature battery electrochemical measurements. **f**, Photographs of 1.0 M AlCl₃-based SOCl₂ electrolyte used for low temperature Li/KJCO₂ battery after being frozen in a -80 °C freezer overnight. **g**, Lighting up an LED using a -80 °C Li/KJCO₂ battery coin cell (CR2032). The lighted LED outside the -80 °C freezer was connected by wires to the battery inside the -80 °C freezer. The loading of KJCO₂ was ~ 2.0 mg cm⁻² in all batteries in this figure.

## 2.3. Spectroscopic investigation of battery reactions on the positive electrode side

For batteries after cycling at -40 °C, SEM imaging showed the formation of 200 - 300 nm LiCl on $KJCO_2$ (Supplementary Fig. 8a) corresponding to reduction of trapped $Cl_2$ to LiCl[20, 21]. Upon charging from 1200 to 5000 mAh g$^{-1}$, the deposited LiCl were increasingly removed/oxidized (SEM in Supplementary Fig. 8b, c and EDX mapping in Supplementary Fig. 8d) accompanied by decreases in the battery electrochemical impedance (Supplementary Fig. 9). The formation of $Cl_2$ (2 LiCl → 2 Li$^+$ + $Cl_2$ + 2 e$^-$) constituted the main charging capacity[20, 21], and reduction of trapped $Cl_2$ in $KJCO_2$ constituted the main discharge plateau, giving the main charge/discharge 4.01 V/3.3 V plateaus (Fig. 3e) with an overall battery reaction of Li + 1/2 $Cl_2$ ↔ LiCl at - 40 °C (see Schematic in Fig. 5a).

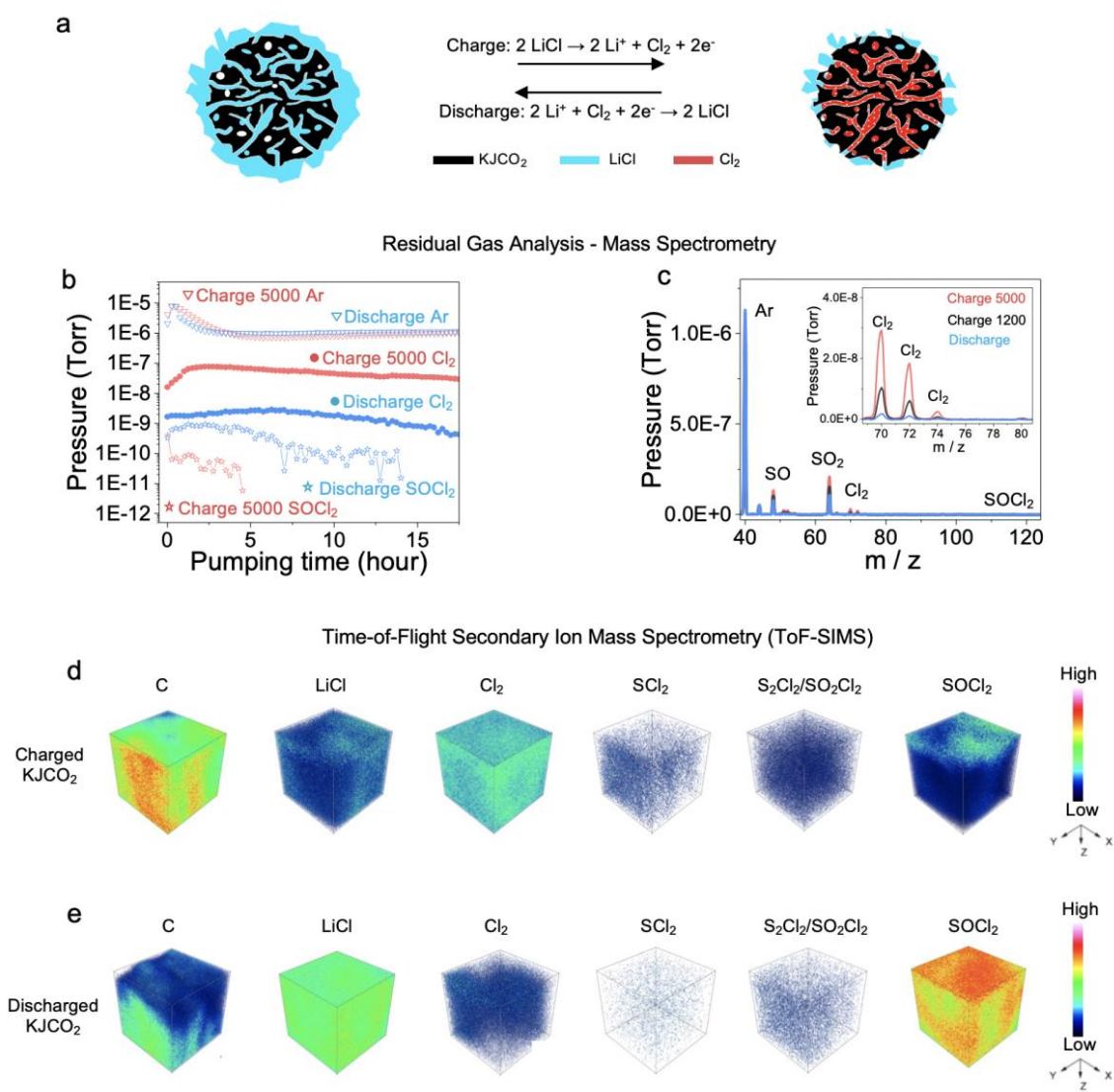

**Figure 5. Battery reactions on the positive electrode side. a**, Schematic drawing of the LiCl oxidation/$Cl_2$ reduction in $KJCO_2$ during charge and discharge, based on the reversible LiCl/$Cl_2$ redox reactions. **b**, The mass spectrometry detected pressure vs. vacuum pumping times of Ar, $Cl_2$, and $SOCl_2$ evolved from a $KJCO_2$ electrode charged to 5000 mAh g$^{-1}$ and those evolved from a discharged cell. Both cells were cycling at -40 °C when stopped for removing the positive electrodes

for measurements. **c.** Mass spectrometry (after 17.5 hours vacuum pumping) of evolved species from $KJCO_2$ electrodes charged to 1200 or 5000 mAh $g^{-1}$ and from discharged ones. Ar (*m/z* = 40 amu), $Cl_2$ (*m/z* = 70 amu), $SOCl_2$ (*m/z* = 118 amu). Inset shows $Cl_2$ with isotopes resolved. **d, e**, 3D distributions (Analysis area: 50 × 50 $\mu m^2$) of C, LiCl, $Cl_2$, $SCl_2$, $S_2Cl_2/SO_2Cl_2$ and $SOCl_2$ secondary ion fragments constructed from TOF-SIMS depth scan of a charged 5000 mAh $g^{-1}$ $KJCO_2$ electrode (d) and fully discharged $KJCO_2$ electrode (e) evolved from -40 °C Li/$KJCO_2$ batteries.

SEM showed that not all of the surface LiCl coating was oxidizable to $Cl_2$ for subsequent reduction and battery cycling (Supplementary Fig. 8c), as observed previously[20]. When cycling/charging to higher capacities (i.e. 5000 mAh $g^{-1}$), LiCl in the carbon electrode was increasingly oxidized to $Cl_2$ accompanied by oxidation of $SOCl_2$ in electrolyte on carbon sites free of salt coating to form $SCl_2/S_2Cl_2$ and $SO_2Cl_2$ (see Fig. 3e and Fig. 4d, the increase in charging voltage, and Supplementary Text 1)[20, 21]. While most of the $Cl_2$ were trapped in the porous carbon for reversible reduction to LiCl in discharge (shown in battery retention experiments below)[20], we postulate that the more reactive $SCl_2/S_2Cl_2$ and $SO_2Cl_2$ species were poorly trapped, migrated into the electrolyte and shuttled to the Li anode side to cause chemical oxidation and corrosion. Such shuttling could lead to anode degradation, consumption of electrolyte that prevented 100% $SOCl_2$ regeneration and decreased battery cycling life[20, 36].

We employed mass spectrometry to probe ex situ trapped species in $KJCO_2$ after cycling at - 40 °C by placing the electrodes under vacuum pumping for mass/charge (*m/z*) sampling of gaseous species evolved (see Methods, Supplementary Fig. 10 and Supplementary Text 2). The detected $Cl_2$ escaping from a charged $KJCO_2$ electrode (to 5000 mAh $g^{-1}$ or 1200 mAh $g^{-1}$) far exceed that from a discharged $KJCO_2$ (Fig. 5b, c, Supplementary Fig. 11). The much lower $Cl_2$ detected in the latter was attributed to fragmentation of $SOCl_2$ in the residual electrolyte[20,21].

Further, we performed 3D chemical mapping by time of flight - secondary ion mass spectroscopy (TOF-SIMS) depth profiling imaging of charged $KJCO_2$ (to 5000 mAh $g^{-1}$ at -40 °C) and discharged $KJCO_2$ in 50 × 50 $\mu m^2$ by area regions (see Methods). In charged $KJCO_2$, we detected low counts of LiCl and high counts of $Cl_2/SCl_2/S_2Cl_2/SO_2Cl_2/SCl_2$/carbon due to electro-oxidation of LiCl and $SOCl_2$ to $Cl_2$, $SCl_2$, $S_2Cl_2$ and $SO_2Cl_2$ and more exposed carbon surfaces free of LiCl coating upon charging (Fig. 5d). These species reduced while LiCl and $SOCl_2$ counts increased upon discharging (Fig. 5e), confirming LiCl formation and regeneration of $SOCl_2$ (Supplementary Text 1 and Text 2).

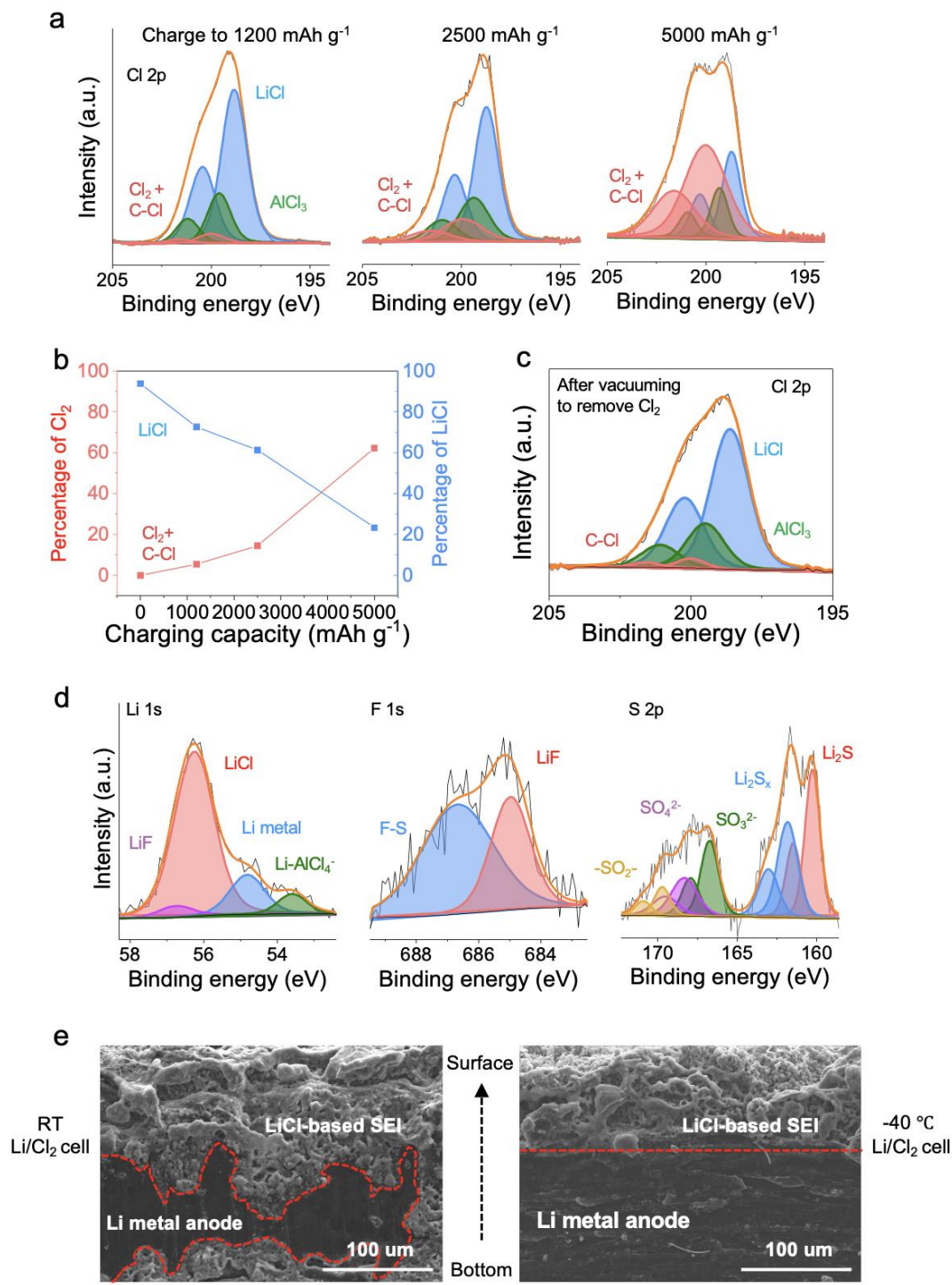

**Figure 6. Battery reactions on the positive and negative electrode side. a,** Cl 2p spectrum of KJCO$_2$ electrode after the -40 °C Li/KJCO$_2$ battery was charged to 1200, 2500, and 5000 mAh g$^{-1}$. Notice the red shaded components increase under increased charging, due to increased Cl$_2$ generated and trapped in KJCO$_2$. **b,** Percentages of Cl$_2$/C-Cl and LiCl species detected in charged KJCO$_2$ to different capacities analyzed from XPS fitting spectra (a). Increased charging led to clear decreases of LiCl and increases of Cl$_2$ detected. **c,** Cl 2p XPS spectrum of the charged 5000 mAh g$^{-1}$ KJCO$_2$ electrode after pumping in vacuum for ~ 50 hours, showing much reduced Cl$_2$ species (in red shaded region under the curve), suggesting removal of trapped Cl$_2$ in KJCO$_2$. **d,** High-resolution XPS spectrum for Li 1s, F 1s and S 2p of the Li metal surface in a -

40 °C Li/KJCO$_2$ battery after 10 cycles, respectively. **e**, Cross-sectional SEM images of Li metal anode in a Li/KJCO$_2$ battery after 10 cycles at room temperature (left) and -40 °C (right) with 100 mA g$^{-1}$ current, respectively.

Lastly, we employed ex-situ XPS to probe species residing in KJCO$_2$ after charged to or discharged from 1200, 2500 and 5000 mAh g$^{-1}$, respectively (within hours of removal of electrodes from batteries, see Methods and Supplementary Text 3) (reference binding energies for the compounds below can be found in Supplementary table 2). In all charged KJCO$_2$ we detected a peak in the Cl 2p spectrum at ~ 200 eV binding energy (Fig. 6a, red lines) but not in discharged KJCO$_2$ (Supplementary Fig. 12). We attributed the ~ 200 eV peak to Cl$_2$ remaining trapped in the KJCO$_2$ and a small percentage of C-Cl, similar to our recent finding with Na/Cl$_2$ batteries[37]. The Cl$_2$ percentage increased with increasing charging capacity from 1200 to 5000 mAh g$^{-1}$, accompanied by decreases in the LiCl XPS signal on KJCO$_2$ (Fig. 6a, b). Note that the same ~ 200 eV peak was reported for adsorbed molecular Cl$_2$ on Ag/AgCl at 100 K[38].

The ~ 200 eV peak in the 5000 mAh g$^{-1}$ charged KJCO$_2$ electrode significantly decreased (Fig. 6c) after ~ 50 hours vacuum pumping (over which mass spectrometry detected little Cl$_2$ remaining in the electrode, Supplementary Fig. 13), indicating that Cl$_2$ remained in charged carbon electrode for up to ~ 2 days under vacuum. Afterwards a weak peak at ~ 200 eV peak remained (Supplementary Fig. 14a) and was attributed to C-Cl formed by chlorination of carbon during battery charging[39]. The C-Cl peak persisted after washing by deionized ultra-filtered (DIUF) water to remove all the residual LiCl, AlCl$_3$ (Supplementary Fig. 14b, c), but disappeared upon annealing at 600 °C in a N$_2$ atmosphere due to C-Cl bond breaking (Supplementary Fig. 14d). With charged KJCO$_2$ we also observed a peak at ~ 286.6 eV in the C 1s spectrum due to Cl bonding (Supplementary Fig. 15)[39].

**2.4. Trapping of reactive species in porous carbon and battery reversibility: temperature effects**

Cycling of our Li/KJCO$_2$ batteries at higher specific capacities generated larger amounts of Cl$_2$, SCl$_2$/S$_2$Cl$_2$ and SO$_2$Cl$_2$ in the positive electrode side during charging (Supplementary Text 1). These species were highly reactive and could shuttle to Li anode to damage the SEI and corrode the underlying Li metal. When such side reaction was severe, higher than 100% CE was observed suggesting that the Li/SEI layer equilibrium was broken by the shuttled species, and the Li metal became more exposed and could discharge beyond the deposited Li by charging (room temperature et al., Supplementary Fig. 5). For batteries stably cycling with ~ 100 % average CE over > 100 cycles, such as Li/KJCO$_2$ cells cycling at 1200 mAh g$^{-1}$ at -40 °C (Fig. 3a), reactive species were effectively trapped in the carbon positive electrodes with little shuttling effect. This was corroborated with battery retention result that when we held a battery charged to 1200 mAh g$^{-1}$ at - 40 °C at open-circuit for 3 days and then discharged the battery, the battery retained ~ 100% CE without any capacity loss (Supplementary Fig. 16a). The retention experiment was repeated several times during cycling over > 100 cycles (Supplementary Fig. 16a).

When a Li/KJCO$_2$ battery was charged at -40 °C and warmed to room temperature to hold for 3 days at open circuit, we observed that the main discharge plateau at ~ 3.3 V decreased in capacity while the lower discharge plateau at ~ 3.10 V extended, accompanied by reduced battery cycle life to < 30 cycles (Supplementary Fig. 16b). The ~ 3.10 V plateau was attributed to the reduction of SO$_2$Cl$_2$ formed by reactions between SO$_2$ and escaped Cl$_2$, due to reduced Cl$_2$ trapping by KJCO$_2$ at higher temperature[20]. A high CE of > 105 % following the retention again indicated shuttling effect.

Temperature dependent battery characteristics suggested that trapping of reactive species in the porous carbon was more effective at lower temperature, preventing shuttling, allowing higher reversible cycling capacity and ~ 100% regeneration of SOCl$_2$ (consumed in the charging step) via LiCl + AlCl$_4^-$ SOCl$^+$ → LiAlCl$_4$ + SOCl$_2$ (Supplementary Text 1)[20]. Complete trapping/retention of electro-oxidation generated species in the positive electrode is therefore critical to prolong the cycle life of our batteries.

Unique to the battery characteristics cycling at -80 °C was that increasing charging capacity from 1200 mAh g$^{-1}$ to ~ 5000 mAh g$^{-1}$ led to more extended high discharge voltage plateaus at ~ 3.8 V (well above the Cl$_2$ reduction voltage plateau of ~ 3.1 V, Fig. 4d), suggesting efficient trapping and reduction of SCl$_2$/S$_2$Cl$_2$ generated in the charging step at -80 °C. The 3.8 V discharge plateau was nearly non-existent in batteries operated at -40 °C (1200 to 5000 mAh g$^{-1}$, Fig. 3e), suggesting escaping of these species from the positive electrodes without subsequent electro-reduction during discharging.

Taken together, at room temperature Li/Cl$_2$ batteries exhibited reversible Li/Li$^+$ and Cl$^-$/Cl$_2$ redox at capacity of ~ 1200 mAh g$^{-1}$ (discharge voltage ~ 3.5 V) with KJCO$_2$ in the positive electrode effectively trapping Cl$_2$ in its abundant nanoscale pores. At a lower temperature of -40 °C, the reversible capacity limit for Cl$^-$/Cl$_2$ redox (discharge voltage ~ 3.3 V) increased to ~ 2500 mAh g$^{-1}$, owing to increased trapping of Cl$_2$ in KJCO$_2$ at -40 °C. At an even lower -80 °C, the reversible capacity limit for Cl$^-$/Cl$_2$ redox further increased to ~ 4000 mAh g$^{-1}$, and the even more reactive species such as SCl$_2$/S$_2$Cl$_2$ were retained on the positive electrode side, giving a high discharge plateau at ~ 3.8 V with up to 1000 mAh g$^{-1}$ capacity. These batteries all showed longer cycle lives at lower temperatures under lower cycling capacities due to reduced shuttling effects. Hence, developing new materials with higher porosity and strong molecular trapping ability for positive electrodes of Li/Cl$_2$ batteries will be a key to boost the capacity of this type of new battery with prolonged cycling life.

## 2.5. Solid-electrolyte interface at the Li negative electrode

Lastly, we investigated the SEI chemistry on the Li metal negative electrode side by ex-situ XPS characterization after removed from a Li/KJCO$_2$ battery cycling at -40 °C for 10 cycles (total cycle life > 130) at 1200 mAh g$^{-1}$ (reference binding energies for the compounds below can be found in Supplementary table 3). The dominant composition in the SEI on the Li electrode was LiCl formed immediately upon exposure of Li to the electrolyte due to reaction with SOCl$_2$ (Fig. 6d). Existence of LiF in the SEI was identified (Fig. 6d), attributed to reactions between Li with FSI- additive in our electrolyte[20, 40-43]. Similar to Na/Cl$_2$ battery and other alkali metal batteries in general[20, 40-43], the fluoride component imparted higher stability to the SEI and cycling stability of the battery (Supplementary Fig. 17). In addition to LiCl and LiF, the SEI also contained F-S related species (FSI$^-$ decomposition, Fig. 6d), -SO$_2$-, Li$_2$SO$_4$, Li$_2$SO$_3$, Li$_2$S, Li$_2$S$_x$, (Fig. 6d), AlCl$_3$ (Supplementary Fig. 18), and SiO$_2$ (residue from separator, Supplementary Fig. 18)[20, 44-46], suggesting reactions between the Li metal and species in the electrolyte including the reactive species from the positive electrode (Cl$_2$, SCl$_2$, SO$_2$Cl$_2$) shuttled to the Li side[20].

We perform SEM imaging of the cross-sectional Li electrodes and observed SEI layers on top of the Li metal. The SEI layer on Li in a battery after cycling at room temperature was thicker and much more irregular than that formed with -40 °C (Fig. 6e). This was attributed to stronger shuttling effect and more severe side reactions/corrosions of Li due to reactive species escaped from the positive electrode, consistent with the reduced cycling performance of Li/Cl$_2$ batteries (Supplementary Fig. 5) and poorer retention of reactive species at higher temperature (Supplementary Fig. 16).

## 3. Conclusions

We developed a novel activated carbon for rechargeable Li batteries in a newly designed SOCl$_2$ based electrolyte exhibiting low viscosity and high conductivity down to low temperatures. The resulting battery delivered reversible capacity in the range of 1200 - 5000 mAh g$^{-1}$ with useful cycle lives at down to -40 °C and -80 °C. Investigations by XPS and mass spectrometry revealed Cl$_2$ and SCl$_2$ /S$_2$Cl$_2$ generated by oxidation of LiCl and SOCl$_2$ during battery charging and trapped in the pore of activated carbon. Low temperatures enhanced trapping of highly oxidative species, slowed down shuttling effects and impeded rapid thickening of the passivating SEI layer on the Li metal anode. The Li/KJCO$_2$ cells are different from other types of secondary batteries in that lower temperatures afford higher battery capacities/energy densities and longer cycle lives than at RT. Battery reversibility relies on trapping of redox active species in the positive electrode and involves electrolyte reaction and regeneration. Innovations in porous carbon and non-carbon (e.g., metal-organic framework MOF[47]) materials for positive electrodes could further improve such batteries towards wide-spread applications.

## 4. Methods

Materials preparation

Thionyl chloride (SOCl$_2$) was purchased from Spectrum Chemical Mfg. Corp. (TH138-100ML) and used as received. Aluminum chloride (AlCl$_3$, 99%, anhydrous, granular) was obtained from Fluka and used without any further purification. Lithium bis(fluorosulfonyl)imide (LiFSI) was purchased from Tianfu Chemical. Lithium chloride (LiCl) was purchased from Ganfeng Lithium Co., Ltd. Lithium metal (thickness 0.38 mm) was purchased from Sigma-Aldrich (7439-93-2). Ketjenblack (KJ) carbon black was commercially available Ketjenblack EC-600JD.

The KJCO$_2$ was synthesized by annealing KJ using a horizontal tube furnace in a CO$_2$-flowing environment (flow rate 200 cc min$^{-1}$). The annealing temperature was 1000 °C and the annealing time was 40 minutes. The temperature was increased from room temperature to 1000 °C at a rate of 10 °C min$^{-1}$. After annealing, the system was allowed to cool down naturally. ~ 85% mass loss was recorded after the annealing with the remaining carbon being KJCO$_2$.

For the electrolyte used in this work, 1.0 M of AlCl$_3$ was added into the SOCl$_2$. After the AlCl$_3$ had completely dissolved, 0.05 M LiFSI was then added into the solution. The mixture was then stirred until all the LiFSI was dissolved. 0.95 M of LiCl was then added into the solution and stirred for ~ 30 minutes or until LiCl could no longer be dissolved. Note that a little excess of LiCl was normally added into the solution in this step to make sure that the electrolyte was completely neutralized. After stirring, a small amount of residual LiCl remained in the solution. The supernatant was used as the electrolyte.

Battery making and electrochemical measurements

90% by weight of either KJ or KJCO$_2$ powder and 10% by weight of PTFE (60% aqueous dispersion, FuelCellStore) were mixed in 100% ethanol (Fisher Scientific). The mixture was then sonicated for 2 hours. Nickel (Ni) foam substrates were cut into circular pieces with diameter of 1.5 cm using a disk cutter (MTI, MSK-T-07). The Ni foams were sonicated in 100% ethanol for 15 minutes and then dried at 80 °C oven until all the ethanol evaporated. The Ni foams were then hovered over a hot plate using a grill, and the mixture of KJ or KJCO$_2$, PTFE and ethanol was then slowly dropped (180 µL each time) onto these Ni foam substrates. The solvent from the previous drop were completely dried before another drop was added to the Ni foam. The carbon mass loading was 2 and 6 mg cm$^{-2}$. The prepared electrodes were then dried at 80 °C overnight and pressed using a roller, and were ready to be used as the positive electrode in batteries.

All batteries were made inside an argon-filled glovebox. A nail file was used to scratch the Li metal (diameter of 1.55 cm) to remove surface contamination. The scratched Li metal was then pasted onto a coin cell spacer (MTI Corporation) and ready for use as the negative electrode. For the Li batteries, KJ or KJCO$_2$ positive electrode, 2 layers of quartz fiber filters (QR-100, Sterlitech), Li negative electrode, and a piece of coin cell spring (MTI Corporation) was put in the center of a CR2032 coin cell positive case (SS316, MTI corporation), with 150 µL of electrolyte added onto the separators. Lastly, the CR2032 negative electrode case (SS316, MTI corporation) was put on top and the entire battery was sealed using a digital pressure controlled electric crimper (MTI corporation, MSK-160E). After the coin cell was assembled, a layer of silicone sealant (GE) was applied on the coin cell to cover the O-ring to further prevent water and air from leaking into the battery.

The battery was operated using a battery tester (Neware, CT-4008-5V50mA-164-U). The low temperature electrochemical measurements of batteries were tested using a -20 °C freezer (So-Low Environmental Equipment), -40 °C freezer (Solidcold Corporation), and -80 °C freezer (STIRLING, ultracold), respectively. The charging step in battery testing was controlled by setting the charging time depending on the cycling capacity (mAh g$^{-1}$) and current condition (mA g$^{-1}$) (charging time =

cycling capacity/current). The discharging step was controlled by setting a discharge cut-off voltage of 2 V. For the batteries tested at room temperature, -20 °C, and -40 °C, the first discharge and cycling current density was 50 mA g$^{-1}$, and 100 mA g$^{-1}$, respectively based on the carbon mass. For the batteries tested at -80 °C, the first discharging and cycling current density was 12.5 mA g$^{-1}$, and 25 mA g$^{-1}$ respectively.

Characterizations

The morphologies of the samples were characterized using field-emission scanning electron microscopy (FE-SEM, Hitachi S-4800) at an accelerating voltage of 15 kV and emission current of 10 μA. In order to obtain the best characterization of pristine KJ and KJCO$_2$, it was necessary to remove the moisture at 100 °C before characterization. For the cycled KJCO$_2$ electrodes and Li metal electrodes, samples were taped to the SEM sample stage using double-sided carbon adhesive in an argon-filled glovebox, and the electrodes should be firmly fixed to prevent magnetic attraction. When transferring the sample stage to the SEM exchange chamber, it was important to minimize exposure to air to avoid oxidation of the samples.

Raman spectra was measured by using the Horiba Jobin Yvon (Olympus BX41) instrument with Ar$^+$ laser of 532 nm. Before each measurement, a piece of p-type boron doped silicon wafer was used for calibration and the silicon peak was adjusted to 520.7 cm$^{-1}$.

XRD measurement was conducted in the Stanford Earth Environmental Measurements Facility using Rigaku MiniFlex 600 Benchtop X-ray Diffraction System. For the cycled KJCO$_2$ electrodes, the batteries were disassembled and the KJCO$_2$ electrodes were taken out from the batteries in an argon-filled glovebox. The KJCO$_2$ electrodes were sealed inside a designed XRD holder that consisted of a polycarbonate dome and an O-ringed 1-inch Si (311) holder in glovebox without exposure to air. The sample holder was then measured at a scan rate of 2° min$^{-1}$.

The Brunauer–Emmett–Teller (BET) surface area and pore volume of pristine KJ and KJCO$_2$ were measured by a 2020 Accelerated Surface Area and Porosimetry System from Micromeritics. Before each measurement, the appropriate amount of carbon was weighed and degassed at 350 °C. After degassing, the weight of the carbon was measured again for final surface area and porosity analysis.

The ionic conductivities of electrolyte were acquired on a FP30 Mettler Toledo conductivity meter at different temperatures. Before measurement, conductivity calibration standards solution was used for calibration. The sensor was placed in the electrolytes stored at room temperature or in freezers, and pressed READ button to start the measurement. The conductivities of electrolyte for different temperatures was then obtained after the display freezed and the decimal point stopped blinking.

The viscosity determination was measured by a Ubbelohde viscometer (Supplementary Fig. 19). At room temperature 10 mL SOCl$_2$ was first added in viscometer, then sucked the air by using a rubber ball. The rubber ball forced this liquid from the bulb through the capillary to the reservoir. The rubber ball was then removed, and the liquid level started to drop from the reservoir to capillary. The time required for the liquid level to pass through two calibrated marks (A line to B line) shown on the capillary was a measure for viscosity, and set as control group ($\eta_0$). Measurements of the 1.0, 1.8, and 4.0 M $AlCl_3$-based electrolytes were the same as for pure $SOCl_2$. The ratios between the recorded time for electrolytes and for pure $SOCl_2$ ($\eta_0$) was used as an indicator of the viscosity changes of electrolyte. For the viscosity determination of 1.0 M $AlCl_3$-based electrolyte at different temperatures, the 1.0 M $AlCl_3$-based electrolyte at room temperature was first measured and set as control group ($\eta_0$), then the electrolyte was moved to freezers for low temperature viscosity measurements.

The electrochemical impedance spectroscopy (EIS) of the battery was measured using a potentiostat/galvanostat (model CHI 760D, CH Instruments). The working electrode was connected to the $KJCO_2$ positive electrode, and the counter and reference electrodes were connected to the Li negative electrode. The initial voltage of the measurement was set to be the open-circuit potential of the battery at the time of the measurement. The high frequency was $1 \times 10^5$ Hz and the low frequency was 0.01 Hz. The amplitude of the measurement was 0.005 V.

Time-of-flight secondary-ion mass spectrometry (TOF-SIMS) measurements used a TOF-SIMS 5-100 instrument from ION-TOF GmbH. To acquire the 3D distribution secondary ion fragments imaging and depth profiles of $KJCO_2$ electrodes, after the Li/$KJCO_2$ battery had cycled to charged or discharged state, the battery was disassembled inside the argon-filled glovebox, and the $KJCO_2$ electrode was then taken out from the battery and sealed into the Al clad pouch. After that, the pouch was sent out for TOF-SIMS measurement about one month the battery was opened. The $KJCO_2$ electrode was taken out from the pouch (exposed to air for about 30 seconds), and transferred into the intro-chamber of TOF-SIMS instrument for vacuuming (~12 hours). After vacuuming, the $KJCO_2$ electrode was transferred into the analysis chamber, with the pressure of the analysis chamber below $1.1 \times 10^{-9}$ mbar. The depth profiling analysis (anion mode) was performed by the $Bi^{3+}$ primary ion beam (scanning area $50 \times 50$ $\mu m^2$, pixel $128 \times 128$), with 30 keV energy, 0.48 pA beam current, and 45° angle. The sputtering ion beam was Cs, with 2 keV energy and 45° angle, and the sputtering area was $200 \times 200$ $\mu m^2$. After depth sputtering and scanning, the 3D distribution secondary ion fragments imaging was reconstructed.

Mass spectrometry measurements were conducted using an RGA-300 residual gas analyzer (Supplementary Fig. 10). After the battery was cycled to its designated state, the $KJCO_2$ electrodes were taken out from the batteries in an argon-filled glovebox and immediately transferred into a designed sample chamber with valve #1 closed. The chamber was then transferred outside the glovebox and connected the capillary tube of RGA-300 system. Valve #2 was opened for 15 min to remove the residual air in the capillary tube. Valve #1 was then opened, and species inside the $KJCO_2$ electrode were continuously pumped to the RGA-300 detector via the capillary tube. The mass spectrometry vs. time then can be obtained.

XPS measurements were conducted at the Stanford Nano Shared Facilities using the PHI

VersProbe 3 instrument. To conduct XPS on electrodes from battery, the samples were clamped onto the XPS stage in an argon-filled glovebox without exposure to air. After sample preparation, the XPS stage was put into an Argon-filled transfer vessel without exposing to air. The transfer vessel was then transferred into the vacuum chamber of the XPS instrument and degassed for about 2 hours for measurement. All the XPS data was calibrated by C 1s 284.8 eV and analysed and fitted using Casa XPS.

**Supporting Information**

Supporting Information is available from the Wiley Online Library or from the author.

**Acknowledgements**

This work was supported by Deng Family gift. Part of this work was performed at the Stanford Nano Shared Facilities (SNSF), supported by the National Science Foundation under award ECCS-2026822. Peng Liang and Guanzhou Zhu contributed equally to this work.

**Contributions**

P.L., G.Z., and H.D conceived the project, performed the characterizations, analysed the data and wrote the paper. C.-L.H. and Y.-Y.L. performed the KJ activation experiments and SEM measurements. H.S. and B.Y. performed the TOF-SIMS experiments. S.-C.W, J.L., F.W., and B.-J.H interpreted the data.

**Competing interests**

The authors declare no competing interests.

**Rechargeable Li/Cl$_2$ battery down to -80 °C**

Peng Liang, Guanzhou Zhu, Cheng-Liang Huang, Yuan-Yao Li, Hao Sun, Bin Yuan, Shu-Chi Wu, Jiachen Li, Feifei Wang, Bing-Joe Hwang, and Hongjie Dai[*]

**Supplementary Text 1**
**Proposed Reactions During Battery Charge-Discharge:**

Upon charging, the deposited LiCl were increasingly removed/oxidized to $Cl_2$ (Supplementary Fig. 8, Fig. 5 and Fig. 6a) accompanied by decreases in the battery electrochemical impedance (Supplementary Fig. 9). The formation of $Cl_2$ constituted the main charging capacity according to the following reaction[1]: $2\ LiCl \rightarrow 2\ Li^+ + Cl_2 + 2\ e^-$. When cycling/charging to higher capacities (i.e. 5000 mAh g$^{-1}$), LiCl in the carbon electrode was increasingly oxidized to $Cl_2$ accompanied by oxidation of $SOCl_2$ in electrolyte on carbon sites free of salt coating. Possible oxidation reactions were proposed including[1]: $SOCl_2 + AlCl_4^- \rightarrow AlCl_4^- SOCl^+ + ½\ Cl_2 + e^-$, $SOCl_2 + SOCl^+AlCl_4^- + 2\ AlCl_4^- \rightarrow SCl^+AlCl_4^- + SO_2Cl_2 + Cl_2 + 2\ AlCl_3 + 2\ e^-$. The $SCl^+AlCl_4^-$ was essentially a compound formed by $AlCl_3$ complexing with $SCl_2$. Another product $SO_2Cl_2$ was known to form from the chemical reaction between $SO_2$ (formed after the first discharge) and $Cl_2$[1]: $SO_2 + Cl_2 \rightarrow SO_2Cl_2$. After enough $Cl_2$ was formed after charge, S was also known to react with $Cl_2$ to form $SCl_2$, which could be further dissociated into $S_2Cl_2$ and $Cl_2$, according to the following reactions[1]: $S + Cl_2 \rightarrow SCl_2$, $2\ SCl_2 \leftrightarrow S_2Cl_2 + Cl_2$. These oxidation reactions led to the coexistence of $Cl_2$, $SCl_2/S_2Cl_2$, and $SO_2Cl_2$ in the electrolyte when the battery was charged, which was trapped in the $KJCO_2$ electrode. During battery discharge (Fig. 3e, Fig. 4d), all the oxidation/charging products of the $Li/Cl_2$ battery were reversibly reduced, corresponding to[1]: $2\ Li^+ + SCl_2 + 2\ e^- \rightarrow S + 2\ LiCl$ and $2\ Li^+ + S_2Cl_2 + 2\ e^- \rightarrow 2\ S + 2\ LiCl$ (~ 4.0 V at -40 °C, ~ 3.8 V at -80 °C), $2\ Li^+ + Cl_2 + 2e^- \rightarrow 2\ LiCl$ (~ 3.3 V at -40 °C, ~ 3.1 V at -80 °C), $2\ Li^+ + SO_2Cl_2 + 2e^- \rightarrow SO_2 + 2\ LiCl$ (~ 3.1 V at -40 °C, ~ 2.9 V at -80 °C)[20, 29]. Low temperature afforded higher reversible cycling capacity of the $Li/Cl_2$ batteries (near 100% CE), with the trapped oxidative species fully reduced to LiCl, and allowing ~ 100% regeneration of $SOCl_2$ through[1]: $LiCl + AlCl_4^- SOCl^+ \rightarrow LiAlCl_4 + SOCl_2$.

**Supplementary Text 2**

**Mass Spectroscopy for Probing Various Species in KJCO$_2$:**

**2.1. RGA**

We employed mass spectrometry to probe ex situ the generated/trapped species in KJCO$_2$ upon battery charging at - 40 °C to 1200 mAh g$^{-1}$ (CH1200) and 5000 mAh g$^{-1}$ (CH5000) respectively. The charged KJCO$_2$ were kept in an Ar-filled chamber after removal from cycling batteries, and the species evolving from the electrode were vacuum pumped and analyzed by a mass-spectrometer over time. (see Methods, Supplementary Fig. 10). In all cases the Ar detected were similar (~ 1E-6 torr) and were used as references (Fig 5b). With a charged 5000 mAh g$^{-1}$ KJCO$_2$ electrode, the detected Cl$_2$ pressure were high (1E-8 to 1E-7 torr), suggesting escaping of trapped Cl$_2$ in the pores of the KJCO$_2$ through vacuum pumping (Fig. 5b, c), and well exceeding Cl$_2$ detected in discharged KJCO$_2$ (Fig. 5b, c). The Cl$_2$ detected in discharged KJCO$_2$ was attributed to fragments of SOCl$_2$ in the electrolyte[1]. Similar trend was also observed in the charged 1200 mAh g$^{-1}$ KJCO$_2$ electrode, but with lower Cl$_2$ pressure (~ 1E-8 torr) than the 5000 mAh g$^{-1}$ case (Fig. 5b, c, Supplementary Fig. 11)

**2.2. 3D Time of Flight - Secondary Ion Mass Spectroscopy (TOF-SIMS) Mapping:**

We performed TOF-SIMS depth scan to determine the distribution of C ($m/z$ = 12), LiCl ($m/z$ = 42), Cl$_2$ ($m/z$ = 70), SCl$_2$ ($m/z$ = 102), S$_2$Cl$_2$/SO$_2$Cl$_2$ ($m/z$ = 134) and SOCl$_2$ ($m/z$ = 118) secondary ion fragments in KJCO$_2$ after charging to 5000 mAh g$^{-1}$ or discharging (50 × 50 μm$^2$ analysis area). In the following discussion, it must be noted that during the TOF-SIMS measurements, parts of detected secondary ions might not originally exist in in KJCO$_2$, but formed as a result of the collision cascade initiated by the highly energetic analysis beam. To investigate the spatial fragment distribution, the TOF-SIMS depth profiles were reconstructed in 3D, where the brighter area reflected the larger counts for various secondary ion fragments (Fig. 5d and 5e). For the charged KJCO$_2$, larger Cl$_2$ and less LiCl counts from the 3D distribution imaging (Fig. 5d) were observed compared to the discharged case (Fig. 5e), due to the oxidation/removal of LiCl to Cl$_2$, accompanied by larger C counts in KJCO$_2$ after charging (Fig. 5d), consistent with the SEM and XPS results. Apart from the Cl$_2$, minute SCl$_2$ and S$_2$Cl$_2$/SO$_2$Cl$_2$ were also observed in the charged KJCO$_2$ due to electrolyte oxidation (Fig. 5d), and its counts decreased in discharged KJCO$_2$ (Fig. 5e), respectively, attributed to the reduction of SCl$_2$ and S$_2$Cl$_2$/SO$_2$Cl$_2$ to LiCl (Supplementary Text 1). These reduction reactions allowed the regeneration of SOCl$_2$ (Supplementary Text 1), which also confirmed by the larger SOCl$_2$ counts in discharged KJCO$_2$ compared to the charged case. (Fig. 5d and 5e).

**Supplementary Text 3**
**Ex-situ XPS for Probing Various Species in KJCO$_2$:**

We employed ex-situ XPS (reference binding energies in Supplementary table 2) to probe various species in KJCO$_2$ after charging to or discharging from 1200, 2500 and 5000 mAh g$^{-1}$, respectively (for electrodes transferred to XPS vacuum within hours after removal from batteries and without any exposing to air, see Methods). Interestingly, apart from the three spin-orbit split doublets (2p$_{3/2}$ and 2p$_{1/2}$; $\Delta_{split}$ = 1.6 eV) of Cl in LiCl and AlCl$_3$ (Fig. 6a)[3, 4], a pronounced peak located at ~ 200 eV was observed in all charged electrodes (Fig. 6a, red lines) and absent in discharged KJCO$_2$ (Supplementary Fig. 12). For the charged samples, we attributed the ~ 200 eV peak to Cl$_2$ remaining trapped in the KJCO$_2$ and C-Cl due to a small fraction of chlorinated carbon. Similar Cl$_2$ peak at ~ 200 eV was observed for adsorbed molecular Cl$_2$ on Ag/AgCl surface at 100 K[2]. The Cl$_2$ percentage from XPS spectral analysis increased with increasing charging capacity from 1200 to 5000 mAh g$^{-1}$, accompanied by decreasing in detected LiCl on KJCl$_2$ surface (Fig. 6a, b). These results corroborated with mass spectrometry data (Fig 5b-e) of oxidation of LiCl to Cl$_2$ and trapping in the electrode upon charging (see schematic in Fig. 5a).

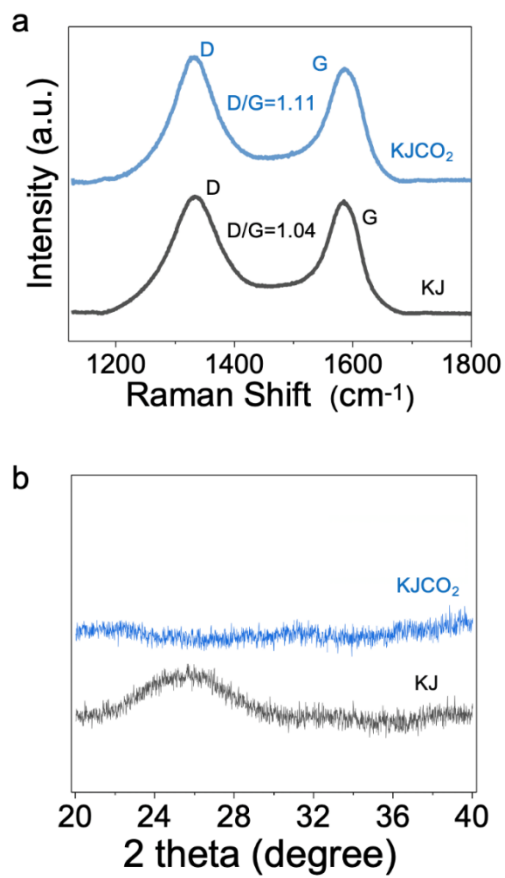

Supplementary Fig. 1. a. Raman spectroscopy of pristine KJ and KJCO$_2$. b. XRD of pristine KJ and KJCO$_2$. Raman spectroscopy showed increased defects and disorder in KJCO$_2$, with a complete disappearance of graphitic ordering in X-ray diffraction after the CO$_2$ activation step.

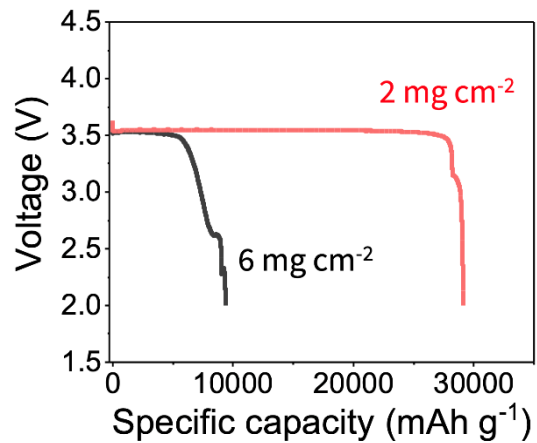

Supplementary Fig. 2. The first discharge curves of room temperature Li/KJCO$_2$ batteries with a KJCO$_2$ loading of ~ 2 and ~ 6 mg cm$^{-2}$. Higher KJCO$_2$ mass loading of ~ 6 mg cm$^{-2}$ still afforded an impressive first discharge capacity of ~ 9,407 mAh g$^{-1}$.

|  | Surface area (m$^2$ g$^{-1}$) | Pore volume (cm$^3$ g$^{-1}$) | Micro pore (cm$^3$ g$^{-1}$) | Meso pore (cm$^3$ g$^{-1}$) |
|---|---|---|---|---|
| KJ | 1307.4 | 3.09 | 0.021 (~ 0.7%) | 3.069 (~ 99.3%) |
| KJCO$_2$ | 2386.9 | 6.53 | 0.036 (~ 0.6%) | 6.494 (99.4%) |

Supplementary Table 1. Average surface area, pore volume (micropores and mesopores) of pristine KJ or KJCO$_2$. The CO$_2$ activation process imparted a ~ 82.6% increase in the specific surface area and a ~ 109.7% increase in the pore volume, allowing for increased deposition of LiCl and higher discharge capacity.

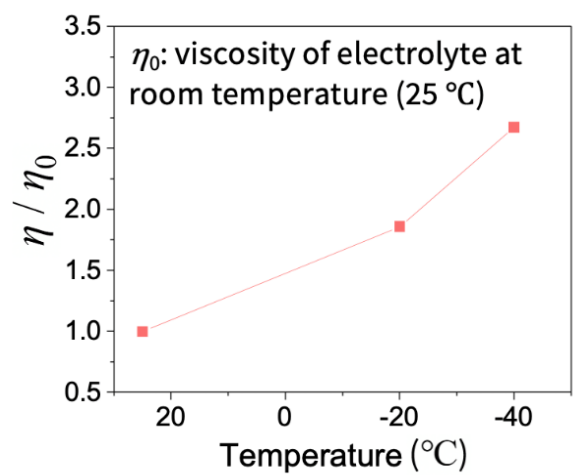

Supplementary Fig. 3. The viscosity of the 1.0 M $AlCl_3$ + 0.95 M LiCl dissolved in $SOCl_2$ with 0.05 M of LiFSI additive electrolyte at room temperature (25 °C), -20 °C, and -40 °C. The viscosity increased as the temperature decreased, leading to the decreasing in discharge capacity.

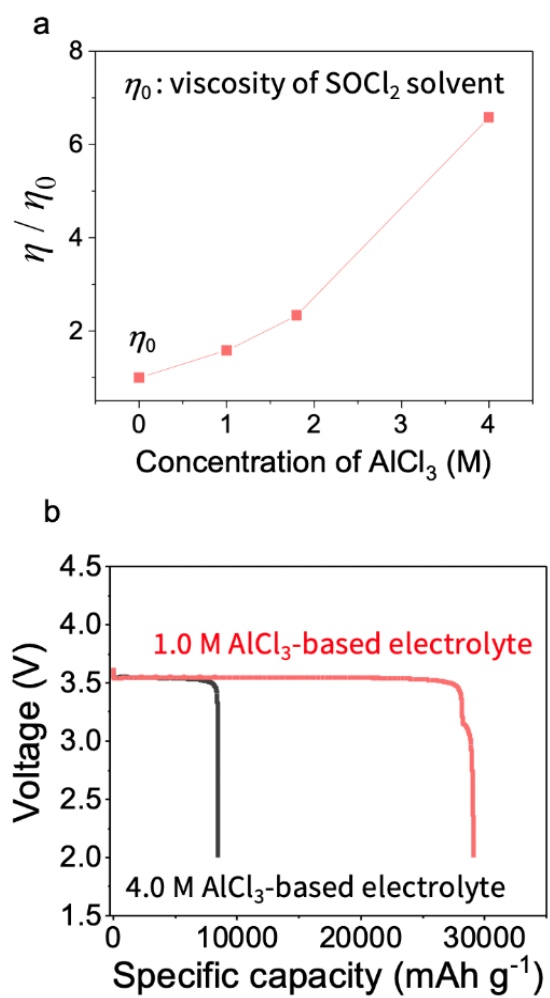

Supplementary Fig. 4. a. The viscosity of the $SOCl_2$ electrolyte with 0 M, 1.0 M, 1.8 M, 4.0 M $AlCl_3$ at room temperatures. b. The first discharge curves of room temperature $Li/KJCO_2$ batteries using $SOCl_2$ electrolyte with 1.0 M and 4.0 M $AlCl_3$. In a 4.0 M $AlCl_3$ based-$SOCl_2$ electrolyte, the battery showed a drastically lower first discharge capacity of 8463 mAh $g^{-1}$ compared to ~ 29,100 mAh $g^{-1}$ in the 1.0 M electrolyte.

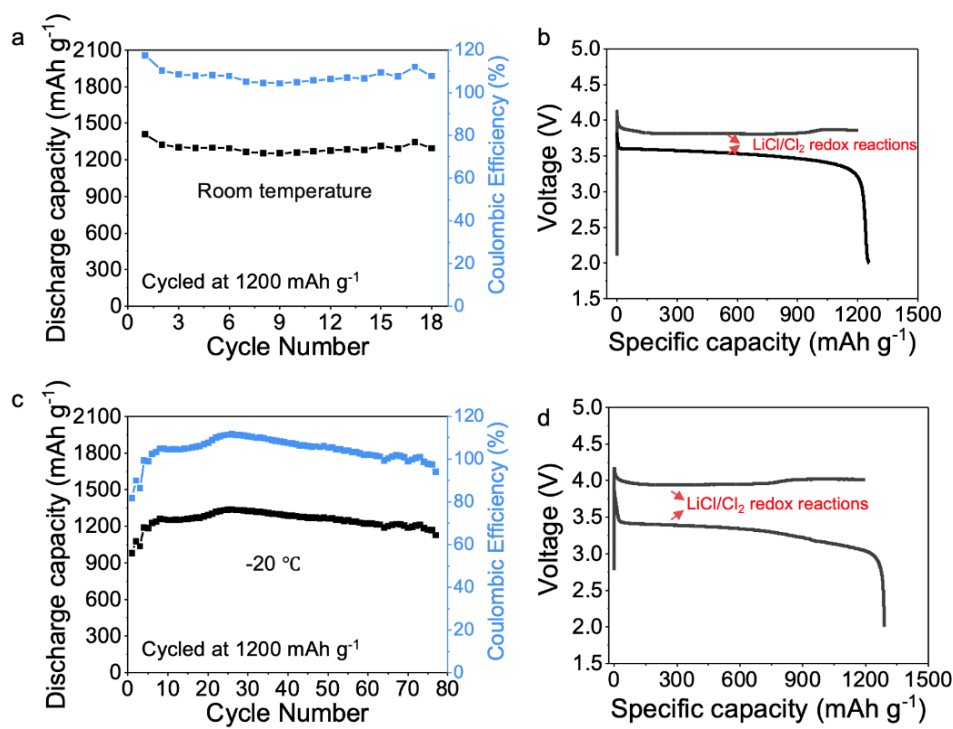

Supplementary Fig. 5. a-b. Cycling performance and discharging/charging curves of the room temperature Li/KJCO$_2$ batteries using KJCO$_2$ at 1200 mAh g$^{-1}$ cycling capacity with 100 mA g$^{-1}$ current. c-d. Cycling performance and discharging/charging curves of the -20 °C Li/KJCO$_2$ batteries using KJCO$_2$ at 1200 mAh g$^{-1}$ cycling capacity with 100 mA g$^{-1}$ current. Rechargeability mainly stemmed from reversible LiCl/Cl$_2$ redox reactions on the positive electrode, i.e., Cl$_2$ was generated by LiCl oxidation during charging and trapped in the pores of KJCO$_2$, and reduced to LiCl upon discharge.

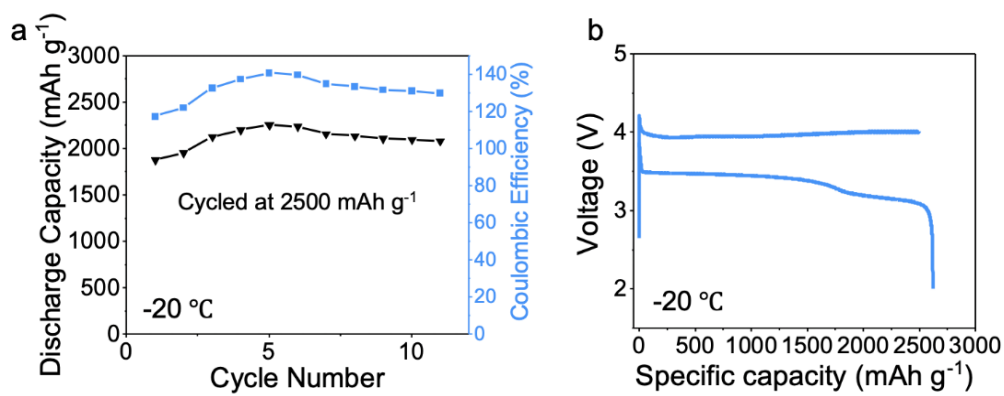

Supplementary Fig. 6. a. Cycling performance of the -20 °C Li/KJCO$_2$ battery at 2500 mAh g$^{-1}$ cycling capacity with 100 mA g$^{-1}$ current. b. Charge-discharge curves of the -20 °C Li/KJCO$_2$ battery at 2500 mAh g$^{-1}$ cycling capacity with 100 mA g$^{-1}$ current.

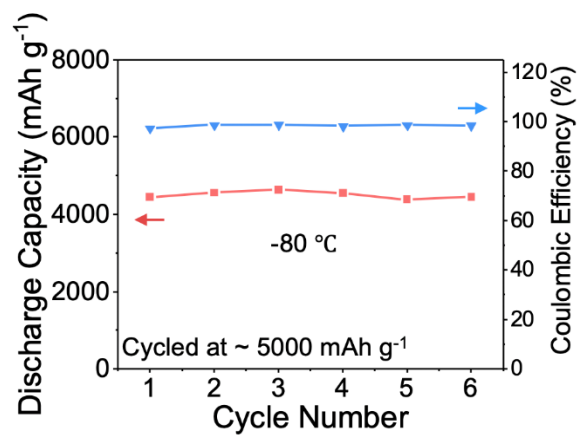

Supplementary Fig. 7. Cycling performance of the -80 °C Li/KJCO$_2$ battery at ~ 5000 mAh g$^{-1}$ cycling capacity with 25 mA g$^{-1}$ current. Note that the voltage limit of battery tester was 5.0 V. Thus, the -80 C battery cannot be charged to 5000 mAh g$^{-1}$ due to the high voltage plateau (reached to 5.0 V) near the end of charging.

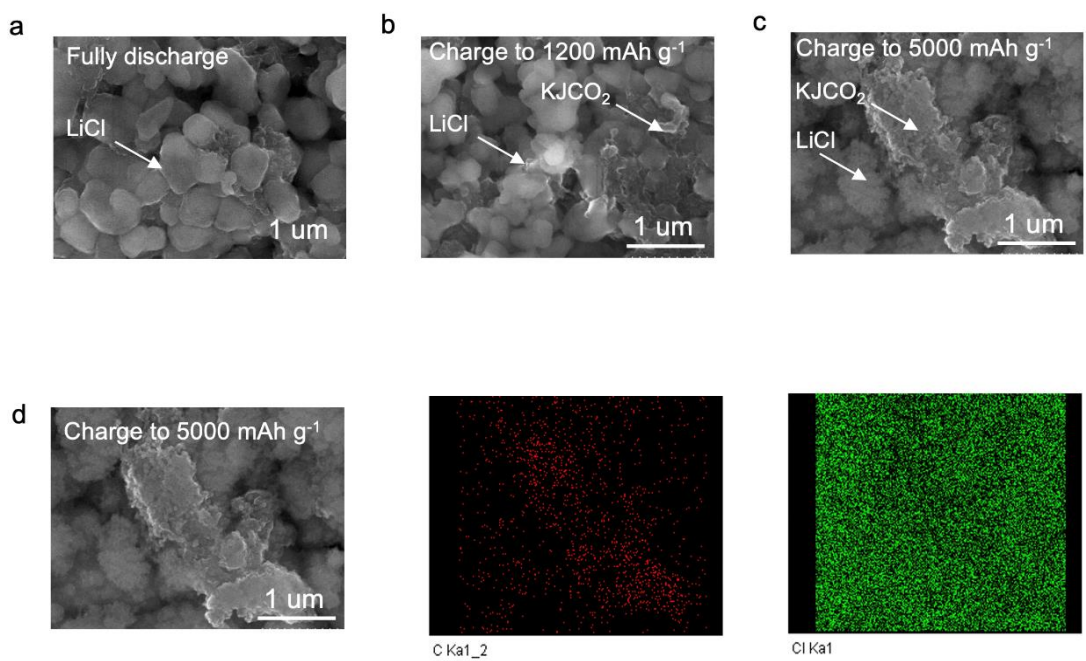

Supplementary Fig. 8. a-c. SEM images of KJCO$_2$ electrode after the -40 °C Li/KJCO$_2$ battery was fully discharged, and charged to 1200 mAh g$^{-1}$ and 5000 mAh g$^{-1}$. d. EDX mapping of KJCO$_2$ electrode after the -40 °C Li/KJCO$_2$ battery was charged to 5000 mAh g$^{-1}$: C signal (red), Cl signal (green). During charging, LiCl on the KJCO$_2$ was oxidized/removed to form Cl$_2$ and the KJCO$_2$ underneath was exposed. However, it was shown that not all of the surface LiCl coating was oxidizable to Cl$_2$ for subsequent reduction and battery cycling. A fraction remained even after charging to 5000 mAh g$^{-1}$ as confirmed by SEM imaging (see the arrow in Supplementary Fig. 8c, the LiCl loosely bound to KJCO$_2$).

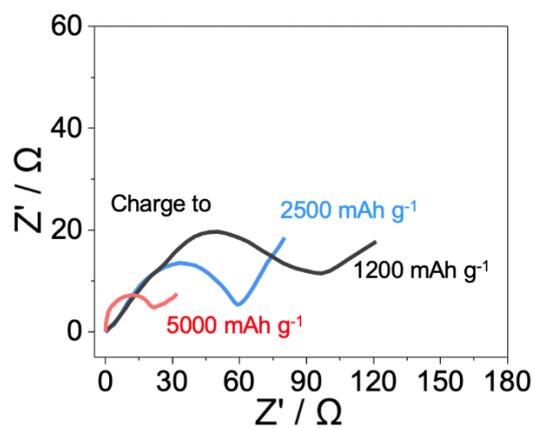

Supplementary Fig. 9. Impedance measurements of the -40 °C Li/KJCO$_2$ battery after charged to 1200 mAh g$^{-1}$, 2500 mAh g$^{-1}$, 5000 mAh g$^{-1}$. The decreases in the battery electrochemical impedance indicated the formation of Cl$_2$ by LiCl oxidation (2 LiCl → 2 Li$^+$ + Cl$_2$ + 2 e$^-$).

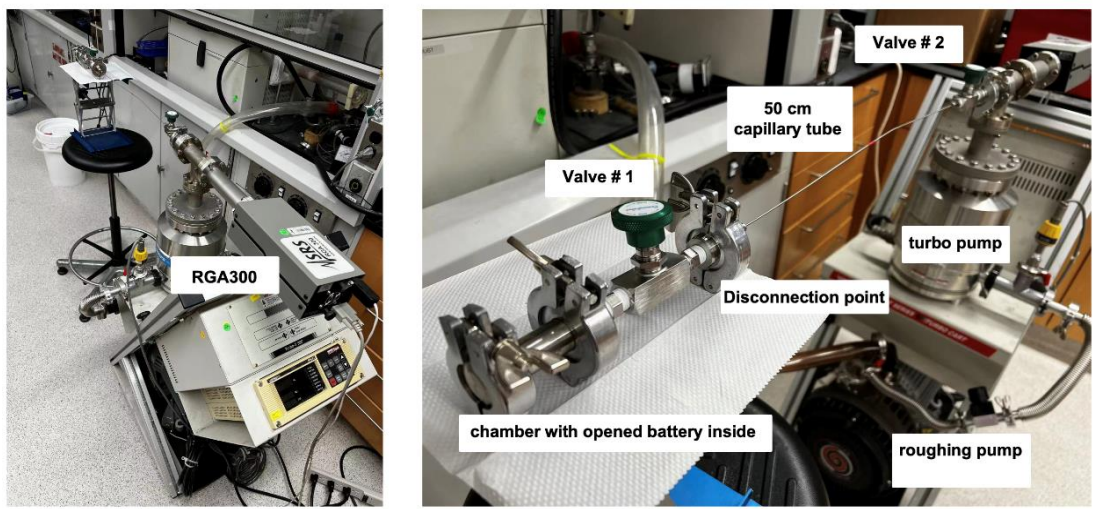

Supplementary Fig. 10. Setup of the RGA-300 (Residual Gas Analyser) instrument for mass spectrometry measurement (See Methods). The charged or discharged $KJCO_2$ were kept in an Ar-filled chamber after removal from cycling batteries, and the species evolving from the electrode were vacuum pumped and analyzed by a mass-spectrometer over time.

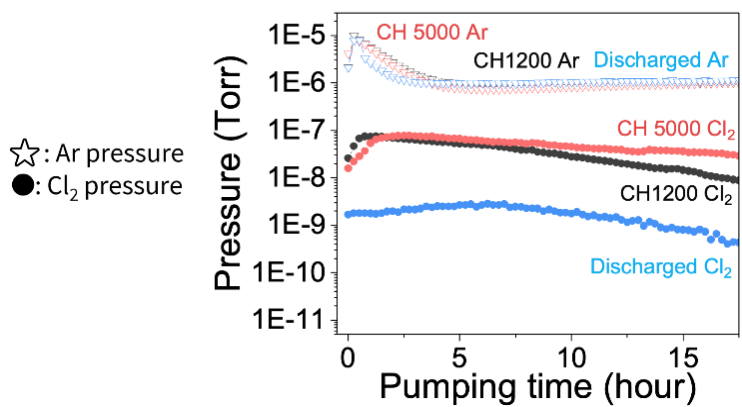

Supplementary Fig. 11. The detected pressure (vs. pumping times) of Ar and $Cl_2$ in $KJCO_2$ electrode of -40 °C Li/$KJCO_2$ batteries after charging to 1200/5000 mAh g$^{-1}$ or discharging. In all cases the Ar detected were similar (~ 1E-6 torr) and were used as references. More $Cl_2$ was detected in $KJCO_2$ electrode when the batteries were charged to higher capacities, suggesting the higher degree of LiCl oxidation to $Cl_2$. The $Cl_2$ detected in discharged $KJCO_2$ was attributed to fragments of $SOCl_2$ in the electrolyte.

| Species | Binding energy / eV | Charge Ref | Ref |
|---|---|---|---|
| Cl2 | 200.0 in Cl 2p | C 1s (284.8 eV) | 2 |
| LiCl | 198.8 in Cl 2p | C 1s (284.8 eV) | 3 |
| AlCl3 | 199.5 in Cl 2p | C 1s (284.8 eV) | 4 |
| C-Cl | 200.0 in Cl 2p | C 1s (284.8 eV) | 5 |
| C-Cl | 286.6 in C 1s | C 1s (284.8 eV) | 5 |
| C-C | 284.8 in C 1s | C 1s (284.8 eV) | 5 |

Supplementary Table 2. Reference binding energies for the compounds shown in Fig. 6a, 6c and Supplementary Fig. 14.

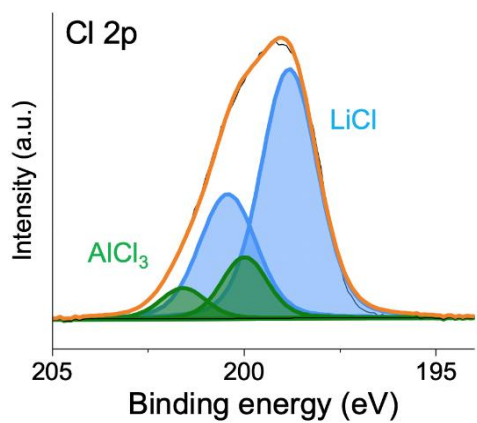

Supplementary Fig. 12. Cl 2p spectrum of KJCO$_2$ after the -40 °C Li/KJCO$_2$ battery was fully discharged. No Cl$_2$ (and C-Cl) peaks in KJCO$_2$ electrode was observed after the battery was discharged, due to the Cl$_2$ (and C-Cl) reduction.

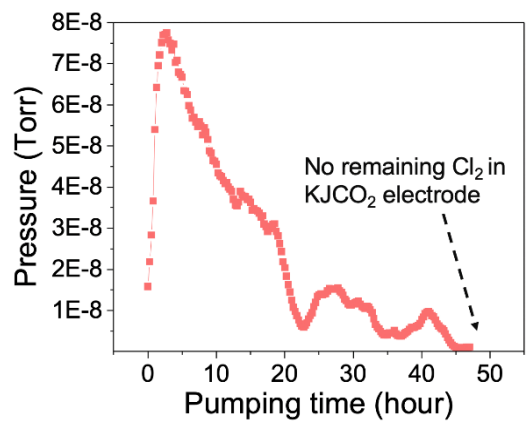

Supplementary Fig. 13. The detected pressure (vs. pumping time) of $Cl_2$ in $KJCO_2$ electrode of a -40 °C Li/$KJCO_2$ battery after charging to 5000 mAh g$^{-1}$ (after ~ 50 hours pumping to remove $Cl_2$).

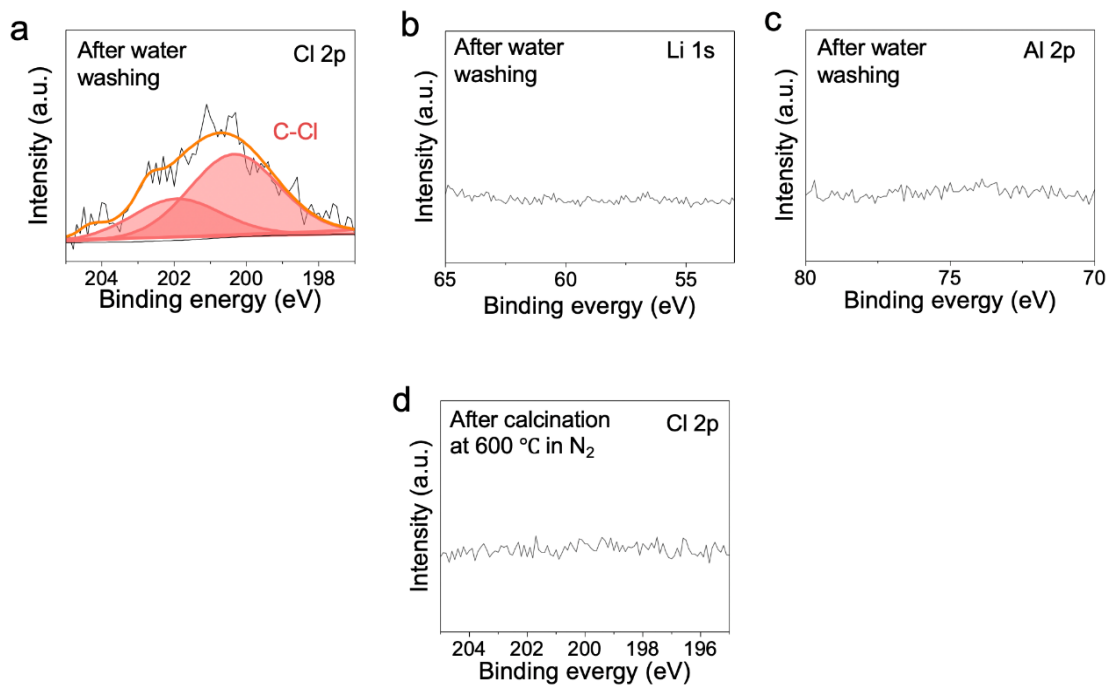

Supplementary Fig. 14. a-c. High resolution Cl 2p, Li 1s and Al 2p XPS spectrum of KJCO$_2$ electrode in the charged 5000 mAh g$^{-1}$ -40 °C Li/KJCO$_2$ battery after washing by deionized ultra-filtered (DIUF) water to remove all the residual LiCl, AlCl$_3$ in electrode (b and c indicated no residual LiCl and AlCl$_3$ in electrode). d. Cl 2p XPS spectrum of KJCO$_2$ electrode in the charged 5000 mAh g$^{-1}$ -40 °C Li/KJCO$_2$ battery after washing and calcination at 600 °C in N$_2$ atmosphere. No peak at ~ 200 eV was observed, suggesting that the C-Cl bonds were broken/removed.

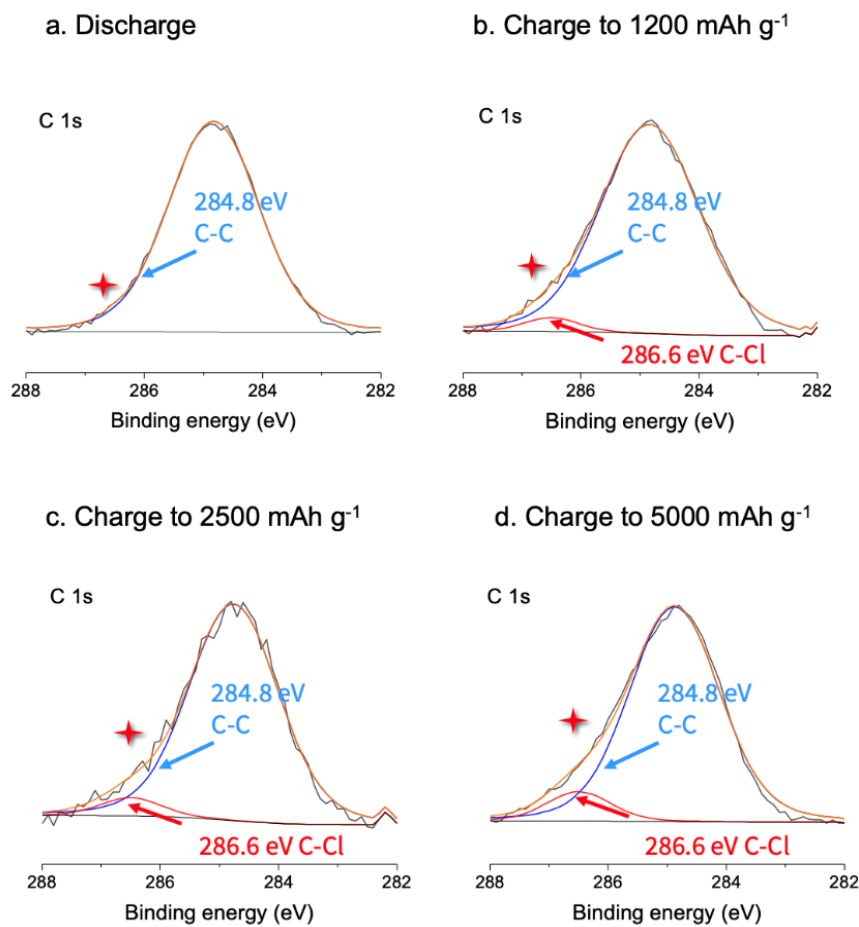

Supplementary Fig. 15. a-d. C 1s spectrum of KJCO$_2$ electrode (without exposing them in Air) in the -40 °C Li/KJCO$_2$ battery after fully discharged (a), charged to 1200 mAh g$^{-1}$ (b), 2500 mAh g$^{-1}$ (c) and 5000 mAh g$^{-1}$ (d). The C-Cl intensity increased as charging capacity increased suggested that more C-Cl bonds formed by chlorination of carbon after the battery was charged to higher capacity, and broken after discharging.

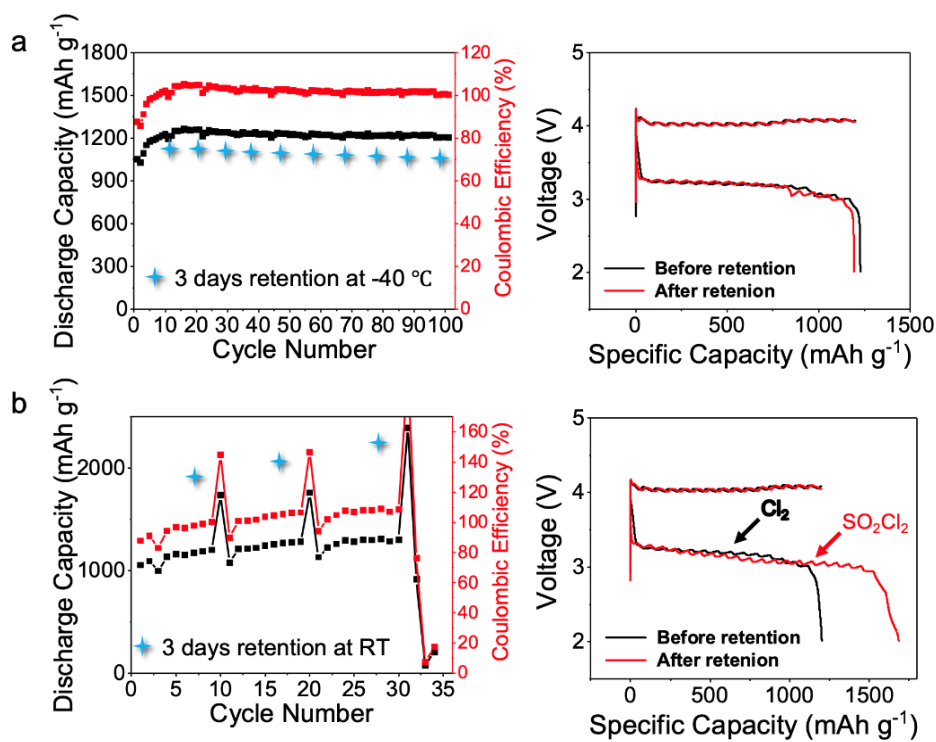

Supplementary Fig. 16. a, b. Cycling performance/curves of the -40 °C Li/KJCO$_2$ batteries using KJCO$_2$ at 1200 mAh g$^{-1}$ after being held in open-circuit for 3 days in charged state at -40 °C (a) and room temperature (b) (operated every 10 cycles, indicated by blue marks).

We held a battery charged to 1200 mAh g$^{-1}$ at - 40 °C at open-circuit for 3 days and then discharged the battery, attaining the same discharge voltage and ~ 100% CE with little charge loss (Supplementary Fig. 16a). The retention experiment was repeated several times during cycling of the battery with a respectable cycle-life of > 100. However, when the battery was charged at -40 °C and moved to room temperature to hold for 3 days at open circuit, we observed that the main discharge plateau at ~ 3.3 V significantly decreased in capacity while the lower discharge plateau at ~ 3.10 V extended, accompanied by reduced battery cycle life to < 30 cycles (Supplementary Fig. 16b). The ~ 3.10 V plateau was attributed to the reduction of SO$_2$Cl$_2$ formed by reactions between SO$_2$ and escaped Cl$_2$, due to lowered Cl$_2$ trapping in KJCO$_2$ at higher temperature.

| Species | Binding energy / eV | Charge Ref | Ref |
|---|---|---|---|
| LiCl | 56.2 in Li 1s<br>198.8 in Cl 2p | C 1s (284.8 eV) | 6 |
| LiF | 56.7 in Li 1s<br>684.8 in F 1s | C 1s (284.8 eV) | 7, 8 |
| Li metal | 54.9 in Li 1s | C 1s (284.8 eV) | 7, 8 |
| Li-AlCl$_4^-$ | 53.6 in Li 1s<br>196.6 in Cl 2p<br>72.5 in Al 2p | C 1s (284.8 eV) | 9 |
| F-S | 686.4 in F 1s | C 1s (284.8 eV) | 10 |
| -SO$_2$- | 169.2 in S 2p | C 1s (284.8 eV) | 10 |
| SO$_4^{2-}$ | 168.7 in S 2p | C 1s (284.8 eV) | 10 |
| SO$_3^{2-}$ | 166.9 in S 2p | C 1s (284.8 eV) | 10 |
| Li$_2$S | 159.8 in S 2p | C 1s (284.8 eV) | 10 |
| Li$_2$S$_x$ | 161.8 in S 2p | C 1s (284.8 eV) | 11 |
| AlCl$_3$ | 199.5 in Cl 2p<br>74.7 in Al 2p | C 1s (284.8 eV) | 9, 12 |
| SiO$_2$ | 532.7 in O 1s | C 1s (284.8 eV) | 13 |

Supplementary Table 3. Reference binding energies for the compounds shown in Fig. 6d and Supplementary Fig. 17.

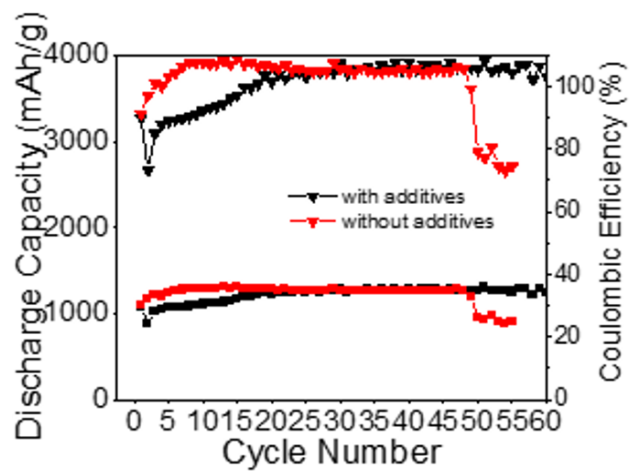

Supplementary Fig. 17. Cycling performance of the -40 °C Li/KJCO$_2$ battery at 1200 mAh g$^{-1}$ cycling capacity using 1.0 M AlCl$_3$-based SOCl$_2$ electrolytes with (black) and without (red) LiFSI additive (100 mA g$^{-1}$ current).

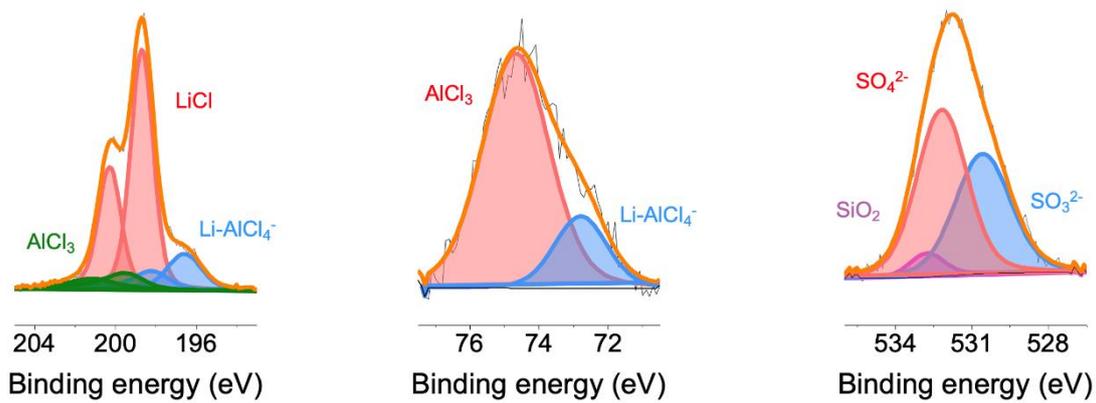

Supplementary Fig. 18. High-resolution XPS spectra for Cl 2p, Al 2p and O 1s of the Li metal surface in a -40 °C Li/KJCO$_2$ battery after 10 cycles, respectively.

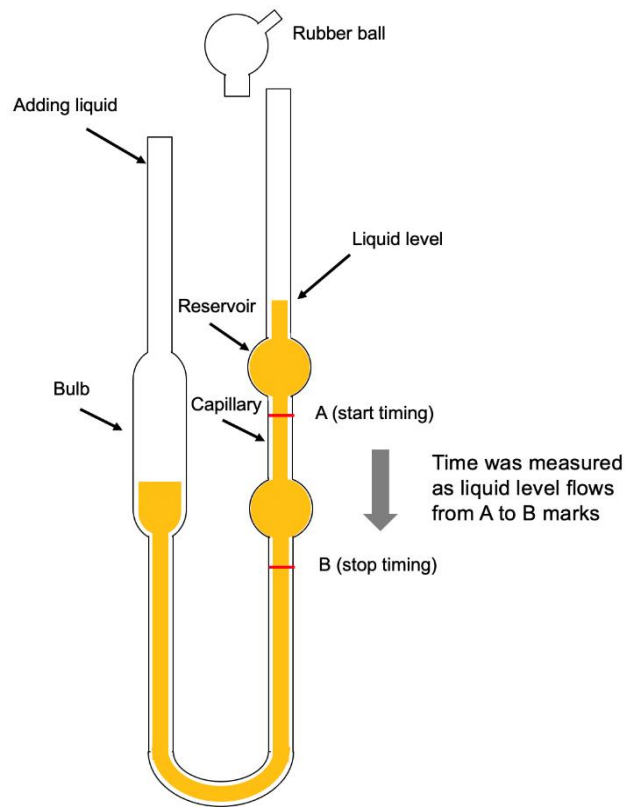

Supplementary Fig. 19. Schematic drawing of the Ubbelohde viscometer (see Methods).